\documentclass[twoside,english]{llncs}
\usepackage[T1]{fontenc}
\usepackage[utf8]{inputenc}
\usepackage[letterpaper]{geometry}
\usepackage{babel}

\usepackage{array}
\usepackage{fancybox}
\usepackage{calc}
\usepackage{amsmath}
\usepackage{graphicx}
\usepackage[unicode=true, pdfusetitle,
 bookmarks=true,bookmarksnumbered=false,bookmarksopen=false,
 breaklinks=false,pdfborder={0 0 1},backref=false,colorlinks=false]
 {hyperref}

\makeatletter

\providecommand{\tabularnewline}{\\}

\newcommand{\threesat}{\textsf{3-SAT}}
\newcommand{\ksat}{\textsf{k-SAT}}
\newcommand{\threecnf}{\textsf{3-CNF}}
\newcommand{\mathematica}{\textsf{Mathematica™}}
\newcommand{\eg}{e.g.}

\usepackage{graphics,graphicx} 
\usepackage{pstricks,pst-node}
\usepackage{wrapfig}
\usepackage{subfigure}

\makeatother

\begin{document}

\title{Non Uniform Selection of Solutions for Upper Bounding the \threesat{}
Threshold%
\thanks{The original publication is available at \protect\href{http://www.springerlink.com}{www.springerlink.com}%
}}

\author{Yacine Boufkhad and Thomas Hugel\\
\email{boufkhad@liafa.jussieu.fr}\\
\email{thomas.hugel@liafa.jussieu.fr}}

\institute{LIAFA - Université Denis Diderot Paris 7 - CNRS%
\thanks{This work was partially supported by GANG project of INRIA.%
}\\
Case 7014\\
F-75205 Paris Cedex 13}
\maketitle
\begin{abstract}
We give a new insight into the upper bounding of the \threesat{}
threshold by the first moment method. The best criteria developed
so far to select the solutions to be counted discriminate among neighboring
solutions on the basis of \emph{uniform} information about each individual
free variable. What we mean by \emph{uniform} information, is information
which does not depend on the solution: e.g. the number of positive/negative
occurrences of the considered variable. What is new in our approach
is that we use \emph{non uniform} information about variables. Thus
we are able to make a more precise tuning, resulting in a slight improvement
on upper bounding the \threesat{} threshold for various models of
formulas defined by their distributions.

\end{abstract}

\section{Introduction}

We consider the phase transition phenomenon that occurs in some random
satisfiability problems, where the probability of satisfiability for
a random formula suddenly goes from $1$ to $0$ at a given ratio
$\frac{\#\mbox{clauses}}{\#\mbox{variables}}$. It was first experimentally
observed that this transition would occur at a ratio near $4.25$
for the standard \threesat{} model (see \cite{Mitchell1992}). The
same kind of transition was also observed in some variants of the
standard model, e.g. when occurrences and signs of variables are balanced
(see \cite{Boufkhad2005}).

The first important step towards the quest of the threshold is the
work of Friedgut and Bourgain \cite{Friedgut1999} establishing that
the width of the transition window tends to zero as the number of
variables tends to infinity.

An important breakthrough was then made by Achlioptas and Peres \cite{Achlioptas2003}:
using a sophisticated technique based on the second moment method
they located asymptotically the threshold of \ksat{} for large constant
$k$ at $2^{k}\ln2-O\left(k\right)$. However in the particular case
of \threesat, there remains a gap between established lower and upper
bounds.

The cornerstone method used for 25 years in order to establish upper
bounds of the \threesat{} threshold is the so called first moment
method. Indeed we are interested in the probability that a formula
has some solutions, but that probability is currently out of reach
of human-tractable calculations; however the moments under this probability
are much easier to estimate. The first moment method consists in bounding
the probability we are interested in by the first moment of a certain
quantity $X$ under this probability. The simplest quantity $X$ one
can imagine as a candidate for the first moment method is the number
of solutions. This gives an upper bound of $5.191$ \cite{ProbabilisticanalysisoftheDavisPutnamprocedureforsolvingthesatisfiabilityproblem::1983::FrancoPaull},
which is far above the experimentally observed threshold at around
$4.25$. There has been ever since lots of efforts \cite{OnRandom3-sat.::1995::MaftouhiVega,Kamath1995,Dubois1997,Kirousis1998,Dubois2000}
intended to lower this upper bound by removing as many solutions as
possible from the counted quantity $X$, the only requirement of the
first moment method being to count at least $1$ solution whenever
a formula is satisfiable; thus the technique is to count only particular
solutions, designed to be present whenever there is a solution, and
not too complicated to count.

We obtain some new upper bounds in a variety of models of \threecnf{}
formulas (which we introduce later in section \ref{sub:Overview-of-Models.}).
In the particular case of the standard model we get an upper bound
of $4.500$. We must mention here the work of Díaz \emph{et al.} \cite{Diaz2009};
gathering the technique of \cite{Dubois2000,Dubois2003} with a pure
literal elimination and a filtering on the typicality of clauses,
they got an upper bound of $4.490$. The fact is that our new technique
is quite compatible with the pure literal elimination and the filtering
on the typicality of clauses, but we only aim at emphasizing the positive
effect of our new technique for selecting solutions, by comparing
it to previous analogous techniques in several models of formulas.

The best implementations of the first moment method approximating
the threshold of \threesat{} use local relationships between solutions,
which involves solutions agreeing on the values of all variables but
a constant number of them, in general one variable \cite{Dubois1997}
or two \cite{Kirousis1998}.

We shall consider the set of solutions with local relationship as
a graph which nodes are the solutions and an edge exists between two
solutions if and only if both solutions agree on the values of all
variables except one. Each edge will be labelled by the variable differing
between both solutions.

\begin{wrapfigure}{r}{50mm}
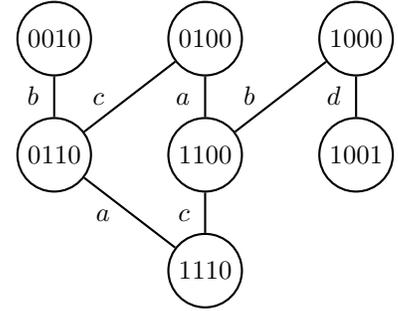

$ \psmatrix[colsep=1cm,rowsep=0.5cm,mnode=circle]
0010&0100&1000\\  0110&1100&1001\\ &1110 \ncline{-}{1,1}{2,1}<{b} \ncline{-}{1,2}{2,1}<{c} \ncline{-}{1,2}{2,2}<{a} \ncline{-}{1,3}{2,2}<{b} \ncline{-}{1,3}{2,3}<{d} \ncline{-}{2,1}{3,2}<{a} \ncline{-}{2,2}{3,2}<{c}
\endpsmatrix $
\caption{Graph of solutions for formula $F$. The label of an edge is the name of the variable differing between both solutions.}   \label{nonoriented} \end{wrapfigure}

For example the formula \begin{eqnarray*}
\Phi & = & \left\{ a\vee b\vee c,a\vee c\vee\overline{d},a\vee\overline{c}\vee\overline{d},a\vee\overline{b}\vee\overline{d},\overline{b}\vee c\vee\overline{d},\overline{a}\vee\overline{b}\vee\overline{d},\overline{a}\vee b\vee\overline{c}\right\} \end{eqnarray*}
 has 7 solutions that can be represented by the non oriented graph
of figure \ref{nonoriented}.

The techniques used so far amount to making an acyclic orientation
of the above graph and to counting only the minimal solutions (those
that do not have outgoing edges). The least is the number of minimal
solutions the best is the upper bound obtained. In general, any graph
can be oriented so as to obtain only one minimal element for every
connected component (e.g. by a depth first search), but this orientation
is obtained thanks to a sophisticated algorithm that is aware of the
whole graph while in our case, the orientation must be decided locally.

\begin{figure}
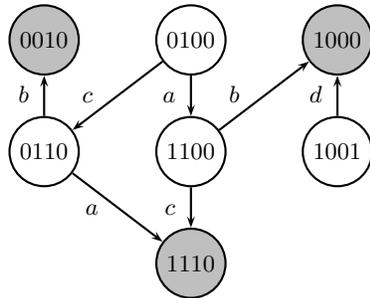
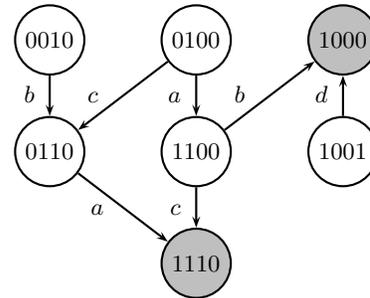

\centering
\subfigure[Uniform orientation. For example $b$ has 2 positive occurrences and 3 negative ones, so every edge labeled by $b$  is oriented from 1 to 0.]{\label{fig:uniform}
$ \psmatrix[colsep=1cm,rowsep=0.5cm,mnode=circle]  [fillstyle=solid,fillcolor=gray!50]0010&0100&[fillstyle=solid,fillcolor=gray!50]1000\\  0110&1100&1001\\ &[fillstyle=solid,fillcolor=gray!50]1110 \ncline{<-}{1,1}{2,1}<{b} \ncline{->}{1,2}{2,1}<{c} \ncline{->}{1,2}{2,2}<{a} \ncline{<-}{1,3}{2,2}<{b} \ncline{<-}{1,3}{2,3}<{d} \ncline{->}{2,1}{3,2}<{a} \ncline{->}{2,2}{3,2}<{c} \endpsmatrix  $}
\hspace{1cm}
\subfigure[Non uniform orientation, obtained in this example by minimizing  $4 \beta_1+2 \beta_2+ \beta_3$ (see definition in section \ref{sub:Notations}). Both edges labeled by $b$ are oriented differently (i.e. from 0 to 1 as well as from 1 to 0).]{\label{fig:nonuniform}
$ \psmatrix[colsep=1cm,rowsep=0.5cm,mnode=circle]  0010&0100&[fillstyle=solid,fillcolor=gray!50]1000\\  0110&1100&1001\\ &[fillstyle=solid,fillcolor=gray!50]1110 \ncline{->}{1,1}{2,1}<{b} \ncline{->}{1,2}{2,1}<{c}	 \ncline{->}{1,2}{2,2}<{a} \ncline{<-}{1,3}{2,2}<{b} \ncline{<-}{1,3}{2,3}<{d} \ncline{->}{2,1}{3,2}<{a} \ncline{->}{2,2}{3,2}<{c} \endpsmatrix  $}
\caption{Two different orientations for the solutions of formula $\Phi$. Minimal solutions are in gray.}
\label{fig:oriented}
\end{figure}

The very first orientation \cite{Dubois1997,Kirousis1998} consisted
in orienting an edge from the solution where the label variable is
assigned $0$ to the one where it is $1$ regardless of which variable
is considered. Later, in \cite{Dubois2003,Diaz2009}, an edge is oriented
towards the value that makes true the most literals and this can be
known thanks to the syntactic property of the number of occurrences
of each variable in the formula. In both these types of orientation,
the edges having the same labels are oriented the same way (e.g. from
$0$ to $1$) anywhere in the graph. So we call such orientations
\emph{uniform }(see Figure \ref{fig:uniform}).

The orientation that we use in this paper is less rigid: two edges
labelled with the same variable can be oriented differently depending
on the solutions involved (that is what we call \emph{non uniform
}orientation, see Figure \ref{fig:nonuniform}). Indeed we keep track
of a set of 5 numbers associated with each variable and use it to
discriminate among neighboring solutions. These 5 numbers provide
information on the repartition of true and false occurrences of each
variable in each type of clauses (clauses having 1, 2 or 3 true literals).
\emph{} Our intuition is that we should select solutions in which
the least occurrences of true literals are \emph{critical}. The less
a clause has true literals, the more its true literals are critical.
Such a property is by nature non uniform.

We develop our technique in a general framework allowing us to apply
it to a wide variety of \threecnf{} models of formulas defined by
their distributions; thus we derive new bounds for some known models
of formulas \cite{Boufkhad2005}. The existence of other non uniform
orientations that may give a smaller number of minimal elements and
then better bounds remains to be investigated. 

In section \ref{sub:Notations} we present our framework and four
different models of formulas; in section \ref{sec:Selection-of-Solutions}
we show how we make our non uniform selection of solutions, and sum
up the bounds we obtain for each model. We give details on the calculation
of the first moment and its constraints in section \ref{sec:The-First-Moment},
as well as some hints on what led us to the weights we took for our
non uniform selection.

\section{Definitions and notations\label{sub:Notations}}

We consider a generic random model of \threecnf{} formulas having
$n$ variables and $cn$ clauses. Models are parametrized by a probability
distribution $\left(d_{p,q}\right)_{p,q\in\boldsymbol{N}}$ such that
$\sum_{p,q\in\boldsymbol{N}}d_{p,q}=3c$. In each model a satisfiability
threshold will appear for a specific value of $c$ we want to estimate.
Before we get formulas we draw \emph{configurations} as follows: 
\begin{enumerate}
\item each of the $n$ variables is given $p$ labelled positive occurrences
and $q$ labelled negative occurrences in a way that the overall proportion
of variables with $p$ positive occurrences and $q$ negatives occurrences
is $d_{p,q}$;
\item a configuration can be seen as a matrix of $3cn$ bins containing
literals occurrences; the repartition of literals into the $3cn$
bins is drawn uniformly among all $\left(3cn\right)!$ permutations
of labelled literals occurrences.
\end{enumerate}
A legal \emph{formula} is a configuration where occurrences are unlabelled
and each clause contains at most one occurrence of each variable.
For the models we consider in this paper and described in section
\ref{sub:Overview-of-Models.}, it was shown that an upper bound on
the satisfiability threshold obtained for configurations also applies
to legal formulas (see \cite{Diaz2009} for the standard model and
\cite{Boufkhad2005} for models where $p$ and $q$ are bounded).
So we shall work on configurations all along this paper.

\subsection{Overview of Models\label{sub:Overview-of-Models.}}
\begin{description}
\item [{Standard~Model:}] all literals are drawn uniformly and independently;
it was shown in \cite{Dubois2003,Diaz2009} that the resulting distribution
is the 2D Poisson distribution: $d_{p,q}={p+q \choose p}\frac{e^{-3c}}{\left(p+q\right)!}\left(\frac{3c}{2}\right)^{p+q}$.
\end{description}
By analogy with the standard model we now define several other models
where we force an equilibrium between variables occurrences and/or
signs. These can be seen as \emph{regular} variants of \threesat{}
(just like regular graphs). The equilibrium cannot be perfect because
of parity or truncation reasons, but we circumvent it as follows.
Of course one can check that all of these distributions sum up to
$1$ and have an average of $3c$.
\begin{description}
\item [{Model~with~Almost~Balanced~Signs:}] every variable appear with
(almost) the same number of positive and negative occurrences; we
define $d_{p,q}$ by $d_{p,p}=\frac{e^{-3c}\left(3c\right)^{2p}}{\left(2p\right)!}$
and $d_{p+1,p}=d_{p,p+1}=\frac{1}{2}\frac{e^{-3c}\left(3c\right)^{2p+1}}{\left(2p+1\right)!}$
(and zero elsewhere).
\item [{Model~with~Almost~Balanced~Occurrences:}] every variable appear
with (almost) the same number occurrences; let $t^{*}=\lfloor3c\rfloor$
and $r^{*}=3c-t^{*}$; we define $d_{p,q}$ by $d_{p,t^{*}-p}=\left(1-r^{*}\right)\frac{{t^{*} \choose p}}{2^{t^{*}}}$
and $d_{p,t^{*}+1-p}=r^{*}\frac{{t^{*}+1 \choose p}}{2^{t^{*}+1}}$
(and zero elsewhere).
\item [{}]~
\item [{Model~with~Almost~Balanced~Signs~and~Occurrences:}] every
variable appear with (almost) the same number occurrences and have
strictly the same number of positive as negative occurrences (this
model was examined in \cite{Boufkhad2005}); let $p^{*}=\lfloor\frac{3c}{2}\rfloor$
and $r^{*}=\frac{3c}{2}-p^{*}$. We define $d_{p,q}$ by $d_{p^{*},p^{*}}=1-r^{*}$
and $d_{p^{*}+1,p^{*}+1}=r^{*}$ (and zero elsewhere).
\end{description}

\subsection{Types of clauses and variables}

Our selection method is based on different types of clauses: given
any assignment, we call clause of \emph{type} $t$ a clause having
$t$ true literals under this assignment, and $\beta_{t}$ the proportion
of clauses of type $t$.

Moreover we want to have some control on the number of occurrences
of variables in the different types of clauses; to do so we need 6
numbers per variable, so we say that a variable is of \emph{type}
$\left(i,j,k,l,m,v\right)$ if it is assigned $v$ and has:

\begin{minipage}[b][1\totalheight][t]{0.5\columnwidth}%
\begin{description}
\item [{i}] true occurrences in clauses of type $1$;
\item [{j}] true occurrences in clauses of type $2$;
\item [{k}] true occurrences in clauses of type $3$;
\item [{l}] false occurrences in clauses of type $1$;
\item [{m}] false occurrences in clauses of type $2$;
\end{description}
\end{minipage}\hfill{}%
\begin{minipage}[t]{0.35\columnwidth}%
\begin{pspicture}(-0.8,0)(2,3) \psframe(0,0)(1.5,2.5)  \psframe(0,0)(1.5,0.5) \psframe[fillstyle=solid,fillcolor=lightgray](1,0.5)(1.5,1.5) \psframe[fillstyle=solid,fillcolor=lightgray](0.5,1.5)(1.5,2.5) \psline(0,1.5)(0.5,1.5) \rput(0.25,2){$i$} \rput(1,2){$l$} \rput(0.5,1){$j$} \rput(1.25,1){$m$} \rput(0.75,0.25){$k$} \rput(1.75,0.25){$\beta_3$} \rput(1.75,1){$\beta_2$} \rput(1.75,2){$\beta_1$} \psline(0.2,1.8)(-0.25,1) \psline(0.2,0.8)(-0.25,1) \psline(0.2,0.3)(-0.25,1) \rput(-0.6,1.2){$true$} \psline(1.3,2)(1.7,2.6) \psline(1.4,1.2)(1.7,2.6) \rput(1.7,2.8){$false$} \end{pspicture}%
\end{minipage}

\begin{remark}
For each variable we have $i+j+k=p$ and $l+m=q$ or vice versa (according
to the value $v$ assigned to the variable). 
\end{remark}
Then we put some weights onto the solutions as follows: in a given
solution each variable of type $\left(i,j,k,l,m,v\right)$ receives
a weight $\omega_{i,j,k,l,m,v}$. The weight of a solution will be
the product of the weights of all variables. It turns out that in
the end we shall take binary weights, yielding in fact an orientation
between solutions. We explain the choice of the weights in sections
\ref{sec:Selection-of-Solutions} and \ref{sub:weights-tuning}. Then
we apply the first moment method to the random variable $X$ equal
to the sum of the weights of the solutions.

\section{Selection of Solutions\label{sec:Selection-of-Solutions}}

Let us recall how the first moment method works: we want to show that
$\mathrm{Pr}\left(Y\geq a\right)$ is small but we don't have access
to $\mathrm{Pr}\left(Y\geq a\right)$. Instead we use some $\mathrm{E}X$.
It suffices then to ensure that $\mathrm{Pr}\left(Y\geq a\right)\leq\mathrm{E}X$.
For our problem \threesat, $Y$ is the number of solutions, $a=1$
and $X$ is the total weight on the solutions. Since $X\geq0$, Markov's
inequality yields that $\mathrm{Pr}\left(X\geq1\right)\leq\mathrm{E}X$;
so if we choose $X$ such that $Y\geq1$ implies $X\geq1$, we have
$\mathrm{Pr}\left(Y\geq1\right)\leq\mathrm{Pr}\left(X\geq1\right)\leq\mathrm{E}X$.
Then our goal will be to tune the weights so that $\mathrm{E}X\to0$
for the least ratio $c=\frac{\#\mbox{clauses}}{\#\mbox{variables}}$.

\subsection{Construction of a Correct Weighting Scheme\label{sub:correction-of-scheme}}

Of course we must put some constraints onto the weights in order that
the weighting scheme can be correct for the first moment method: namely
the sum of the weights of the solutions of a satisfiable formula must
be at least $1$. However the constraints we choose here might not
be necessary for the first moment method to hold.

Let us recall that given a solution, a variable is called \emph{free}
when the assignment obtained by inverting its value (0/1) remains
a solution. Thus in our framework, a variable is free iff its $i$
number is $0$. How does the tuple $\left(0,j,k,l,m,v\right)$ for
a free variable $x$ behave when the value $v$ is inverted to $1-v$?
$i\left(x\right)\leftarrow0$, $j\left(x\right)\leftrightarrow l\left(x\right)$,$k\left(x\right)\leftrightarrow m\left(x\right)$
and $v\left(x\right)\gets1-v\left(x\right)$.
\begin{enumerate}
\item the first constraint we put is that $\omega_{i,j,k,l,m,v}=1$ as soon
as $i\geq1$; that is, we put significant weights only onto free variables.
The reason for this is that free variables allow to move between solutions.
\item the second constraint is that \begin{equation}
\omega_{0,j,k,l,m,v}+\omega_{0,l,m,j,k,1-v}=1\enskip;\label{eq:neighbors}\end{equation}
that is, the sum of the weights of a free variable in a couple of
solutions differing only on that variable is $1$. We impose this
condition by analogy with the conditions on weights given in \cite{Ardila2009a}.

\end{enumerate}
As suggested by the analysis given in section \ref{sub:weights-tuning},
we shall take $\omega_{0,j,k,l,m,v}=\boldsymbol{1}_{P\left(j,k,l,m,v\right)}$
for a certain predicate $P\left(j,k,l,m,v\right)$ linked with the
sign of $\alpha_{1}\rho_{j,l}+\alpha_{3}\rho_{k,m}$ (where $\alpha_{1}$
and $\alpha_{3}$ are any real constants and $\rho$ is an operator
defined as $\rho_{a,b}=a-b$).

The fact that we imposed $\omega_{0,j,k,l,m,v}+\omega_{0,l,m,j,k,1-v}=1$
tells us that given a solution and a free variable $x$ at the value
$v$, the predicate $P$ is satisfied by $x$ at the value $v$ or
(exclusively) by $x$ at the value $1-v$. Thus we are able to define
an orientation between neighboring solutions.

Let us say that variable $x$ is \emph{obedient} when $P$ is satisfied.
We put an arc between 2 solutions differing only on 1 (free) variable
$x$ from the solution $\boldsymbol{S}_{d}$ (where $x$ is disobedient)
to the solution $\boldsymbol{S}_{o}$ (where $x$ is obedient), and
we call that relation $\boldsymbol{S}_{d}>\boldsymbol{S}_{o}$. The
notation $>$ is not randomly chosen.

Namely our weighting scheme counts $1$ for a solution when it does
not have any disobedient free variables, and $0$ otherwise; but what
can ensure that whenever there is a solution, there is also a solution
where all free variables are obedient? It suffices that the relation
$>$ is circuit-free. Then the transitive closure of $>$ is an order,
and we are precisely counting the minimal solutions in that order.
Minimal solutions exist because the set of all solutions is finite.
So let us see how we can make the relation $>$ circuit-free.

\subsubsection{Recapitulation of Existing Methods.}
\begin{description}
\item [{All~Solutions:}] This method consists in computing the first moment
on all solutions: $P\left(j,k,l,m,v\right)\equiv1$.
\item [{Negatively~Prime~Solutions~(NPS):}] This method consists in
counting only solutions which free variables are assigned $1$. That
is $P\left(j,k,l,m,v\right)\equiv v>0$. This method was introduced
in \cite{Dubois1997}.
\item [{NPS~with~Imbalance:}] This method was introduced in \cite{Dubois2003}
and combined to some other ingredients in \cite{Diaz2009}. This method
consists in allowing free variables to take only a value such that
the number of true occurrences is larger than the number of negative
occurrences of this variable (and in case of equality, ties are broken
in favor of the value $1$). In other words $P\left(j,k,l,m,v\right)\equiv\left(\rho_{j,l}+\rho_{k,m},v\right)>_{\mathrm{lex}}\left(0,0\right)$,
where $>_{\mathrm{lex}}$ denotes the lexicographical order.
\end{description}

\subsubsection{Our Method.}

May we choose arbitrary real coefficients $\alpha_{1}$ and $\alpha_{3}$
in the expression of $\alpha_{1}\rho_{j,l}+\alpha_{3}\rho_{k,m}$
in order that the first moment method should hold? It turns out that
it is the case, and here is a proof of it.

We make the following observation: how does the population of the
3 different types of clauses evolve when a free variable $x$ is flipped?
$\beta_{1}+=\rho_{j,l}\left(x\right)$, $\beta_{2}+=\left(\rho_{k,m}-\rho_{j,l}\right)\left(x\right)$
and $\beta_{3}+=-\rho_{k,m}\left(x\right)$.

Thus $\alpha_{1}\rho_{j,l}+\alpha_{3}\rho_{k,m}$ is the variation
of $\alpha_{1}\beta_{1}-\alpha_{3}\beta_{3}$; so we may define our
predicate $P$ in the following way: $P\left(j,k,l,m,v\right)\equiv\left(\alpha_{1}\rho_{j,l}+\alpha_{3}\rho_{k,m},v\right)>_{\mathrm{lex}}\left(0,0\right)$;
thanks to $v$ we break ties when $\alpha_{1}\rho_{j,l}+\alpha_{3}\rho_{k,m}=0$,
so that the underlying relation $>$ between solutions is circuit-free:
namely going from $\boldsymbol{S}_{d}$ to $\boldsymbol{S}_{o}$ when
$\boldsymbol{S}_{d}>\boldsymbol{S}_{o}$ strictly increases $\left(-\alpha_{1}\beta_{1}+\alpha_{3}\beta_{3},v\right)$
for $>_{\mathrm{lex}}$.\\
Moreover the exclusion between $P\left(j,k,l,m,v\right)$ and $P\left(l,m,j,k,1-v\right)$
is satisfied, which means that whenever there is a solution with a
disobedient free variable, it suffices to flip the value of this variable
so that it becomes obedient.\\
We investigated the best ratio between $\alpha_{1}$ and $\alpha_{3}$
by numerical experiments.

\subsection{Summary of Results\label{sub:Summary-of-results}}

\begin{table}
\centering{}\caption{Summary of our results.\label{tab:Summary-of-results-1}}
\begin{tabular}{|>{\centering}p{0.12\paperwidth}|>{\centering}p{0.1\paperwidth}|>{\centering}p{0.1\paperwidth}|>{\centering}p{0.1\paperwidth}|>{\centering}p{0.1\paperwidth}|}
\hline 
model & standard & almost balanced signs & almost balanced occurrences & almost balanced signs and occurrences\tabularnewline
\hline 
all solutions & $5.040$ & $3.858$ & $5.046$ & $3.783$\tabularnewline
\hline 
NPS

$v>0$ & $4.552$ & $3.521$ & $4.662$ & $3.548$\tabularnewline
\hline 
NPS+imbalance

$\left(\rho_{j,l}+\rho_{k,m},v\right)>\left(0,0\right)$ & $4.506$ & $3.514$ & $4.628$ & $3.548$\tabularnewline
\hline 
our method

$\left(\alpha\rho_{j,l}+\rho_{k,m},v\right)>\left(0,0\right)$ & $4.500$ & $3.509$ & $4.623$ & $3.546$\tabularnewline
\hline 
our $\alpha$ & $\alpha=2.00$ & $1.01\leq\alpha\leq1.16$ & $2.01\leq\alpha\leq2.24$ & $\alpha\geq1.01$\tabularnewline
\hline
\end{tabular}
\end{table}

As one can see in table \ref{tab:Summary-of-results-1}, our method
yields in all models a slight improvement on the bounds obtained by
former methods. Note that for some models there is a range of values
for $\alpha$ which give the same upper bound.

In the model where signs as well as occurrences are balanced, the
method of NPS+imbalance is of course the same as the method of NPS,
whereas our method is somewhat better than the method of NPS.

The bound we obtain in the standard model is 4.500; this is not better
than the bound of 4.490 obtained by Díaz \emph{et al.} in \cite{Diaz2009}.
Their calculation adds 2 ingredients to the method of \cite{Dubois2003}:
typicality of clauses and elimination of pure literals. These 2 ingredients
might be combined to our approach to improve on the 4.490, but this
would involve too complicated calculations with respect to the expected
improvement. However in models where signs are balanced it is irrelevant
to eliminate pure literals.

\section{The First Moment Method\label{sec:The-First-Moment}}

\subsection{Types of variables}

We split the set of variables into several sets and subsets of variables.
In order to be able to match the original random \threecnf{} model
of formulas where all literals are drawn independently, we should
consider $p$ and $q$ to range in $\boldsymbol{N}$. For convenience
of our forthcoming maximization, we only take into account bounded
values of $p$ and $q$. So we are going to consider 2 kinds of variables,
according to their numbers of occurrences. We follow the notations
of \cite{Diaz2009}. We denote by $M$ some integer whose value will
be determined according to the required accuracy of the calculations;
in practice we shall take $M=21$. $M$ enables us to define 2 kinds
of variables:
\begin{enumerate}
\item the set of light variables, that is variables which indices are in
the set \begin{eqnarray}
\mathcal{L} & = & \left\{ \left(p,q\right)\in\boldsymbol{N}^{2},p\leq M\land q\leq M\land d_{p,q}>0\right\} \enskip;\end{eqnarray}
they are the most important variables since almost all variables are
light in the models we consider; we call $\delta_{p,q}$ the proportion
of light variables having $p$ positive occurrences, $q$ negative
occurrences, and assigned $1$. As a further refinement, we call $\pi_{i,j,k,l,m,v}$
the proportion of variables of type $\left(i,j,k,l,m,v\right)$ whose
corresponding weight $\omega_{i,j,k,l,m,v}$ is non zero, and omit
the other ones because we shall need all active $\pi_{i,j,k,l,m,v}$
to be non zero. To connect $\pi_{i,j,k,l,m,v}$'s with $\delta_{p,q}$'s
we introduce the following set of tuples of integers: $Q_{p,q}=\left\{ \left(i,j,k,l,m\right)\in\mathbf{N}^{5},i+j+k=p\land l+m=q\right\} $;
thus we have \begin{eqnarray}
\sum_{\substack{\left(i,j,k,l,m\right)\in Q_{p,q}}
}\pi_{i,j,k,l,m,1} & = & \delta_{p,q}\enskip;\label{eq:Qpq}\\
\sum_{\substack{\left(i,j,k,l,m\right)\in Q_{q,p}}
}\pi_{i,j,k,l,m,0} & = & d_{p,q}-\delta_{p,q}\enskip.\label{eq:Qqp}\end{eqnarray}
 Note that equality \ref{eq:Qqp} involves $Q_{q,p}$ whereas equality
\ref{eq:Qpq} involves $Q_{p,q}$.
\item the set of heavy variables, that is all other variables; their indices
are thus in the set \begin{eqnarray}
\mathcal{H} & = & \left\{ \left(p,q\right)\in\boldsymbol{N}^{2},p>M\lor q>M\lor d_{p,q}=0\right\} \enskip;\end{eqnarray}
we weaken the notion of satisfiability by considering that heavy variables
are always satisfied, regardless of their signs and values. Doing
so is harmless for the validity of the first moment method because
we can only increase the number of solutions. In other words we are
going to consider heavy variables as undistinguishable members of
a tote bag. We call $\tau$ the global scaled number of heavy variables:
$\tau=\sum_{\left(p,q\right)\in\mathcal{H}}d_{p,q}$.
\end{enumerate}
We also need to distinguish some types of occurrences of heavy variables.
We call $H$ the global scaled number of occurrences of heavy variables:\\
 $H=\sum_{\left(p,q\right)\in\mathcal{H}}\left(p+q\right)d_{p,q}=3c-\sum_{\left(p,q\right)\in\mathcal{L}}\left(p+q\right)d_{p,q}$.
According to the types of clauses where occurrences appear, $H$ is
divided into $H_{t}$'s, where $H_{t}$ is the scaled number of occurrences
of heavy variables in clauses of type $t$.

We are now ready to write down the expression of the first moment
of $X$, the weight of all solutions.

\subsection{Expression of the First Moment and its Constraints}

We recall that all occurrences of literals are drawn according to
the distribution $d_{p,q}$ (see section \ref{sub:Notations}). Thus
the sample space we consider consists in the $\left(3cn\right)!$
permutations of labelled occurrences of literals, and our parameters
are $n$, $c$, $d_{t,p}$, $\tau$, $H$ and $\omega_{i,j,k,l,m,v}$'s
(although we must carefully choose the weights $\omega_{i,j,k,l,m,v}$,
as explained below in section \ref{sub:weights-tuning}).

All other quantities: $\beta_{t}$, $H_{t}$, $\delta_{t,p}$ and
$\pi_{i,j,k,l,m,v}$ are variables, and the first moment of $X$ can
be split up into a big sum over all variables of the product of the
following factors depending on variables: number of assignments, weight
of an assignment and probability for an assignment to be a solution.
\begin{enumerate}
\item number of assignments: each variable is assigned $0$ or $1$: $2^{\tau n}\prod_{\left(p,q\right)\in\mathcal{L}}{d_{p,q}n \choose \delta_{p,q}n}$;
\item weight of an assignment: $\prod_{\left(p,q\right)\in\mathcal{L}}\prod_{\substack{\left(i,j,k,l,m\right)\in Q_{p,q}\\
v\in\left\{ 0,1\right\} }
}\omega_{i,j,k,l,m,v}^{\pi_{i,j,k,l,m,v}n}$;
\item probability for an assignment to be a solution: quotient of the number
of satisfied configurations by the total number of configurations:

\begin{enumerate}
\item number of satisfied configurations: a configuration can be seen as
a set of bins filled with occurrences of literals:

\begin{enumerate}
\item each of the $3cn$ bins is first given a truth value:\\
there are ${cn \choose \beta_{1}cn,\beta_{2}cn,\beta_{3}cn}3^{\left(\beta_{1}+\beta_{2}\right)cn}$
possibilities, and the following constraint appears:\begin{eqnarray}
\beta_{1}+\beta_{2}+\beta_{3} & = & 1\enskip.\label{eq:bt}\end{eqnarray}

\item each light literal is given a tuple $\left(i,j,k,l,m\right)$ consistently
with $d_{p,q}$ and $\delta_{p,q}$. This gives a series of constraints:\begin{eqnarray}
\sum_{\substack{\left(i,j,k,l,m\right)\in Q_{p,q}}
}\pi_{i,j,k,l,m,1}+\sum_{\substack{\left(i,j,k,l,m\right)\in Q_{q,p}}
}\pi_{i,j,k,l,m,0} & = & d_{p,q}\enskip.\label{eq:dpq}\end{eqnarray}
Note that $\delta_{p,q}=\sum_{\left(i,j,k,l,m\right)\in Q_{p,q}}\pi_{i,j,k,l,m,1}$.
Thus, given a family $\left(\pi_{i,j,k,l,m,v}\right)$, there are
\begin{eqnarray*}
\prod_{\left(p,q\right)\in\mathcal{L}}{\delta_{p,q}n \choose \dots\pi_{i,j,k,l,m,1}n\dots}_{\substack{\left(i,j,k,l,m\right)\in Q_{p,q}}
}\\
\cdot\prod_{\left(p,q\right)\in\mathcal{L}}{\left(d_{p,q}-\delta_{p,q}\right)n \choose \dots\pi_{i,j,k,l,m,0}n\dots}_{\substack{\left(i,j,k,l,m\right)\in Q_{q,p}}
}\end{eqnarray*}
 possible allocations. Moreover the following constraints appear,
so that all occurrences of literals can fit into the destined types
of clauses:\begin{eqnarray}
\sum_{\substack{\left(p,q\right)\in\mathcal{L}\\
\left(i,j,k,l,m\right)\in Q_{p,q}\\
v\in\left\{ 0,1\right\} }
}i\pi_{i,j,k,l,m,v}+H_{1} & = & \beta_{1}c\enskip;\label{eq:i}\\
\sum_{\substack{\left(p,q\right)\in\mathcal{L}\\
\left(i,j,k,l,m\right)\in Q_{p,q}\\
v\in\left\{ 0,1\right\} }
}j\pi_{i,j,k,l,m,v}+H_{2} & = & 2\beta_{2}c\enskip;\label{eq:j}\\
\sum_{\substack{\left(p,q\right)\in\mathcal{L}\\
\left(i,j,k,l,m\right)\in Q_{p,q}\\
v\in\left\{ 0,1\right\} }
}k\pi_{i,j,k,l,m,v}+H_{3} & = & 3\beta_{3}c\enskip;\label{eq:k}\\
\sum_{\substack{\left(p,q\right)\in\mathcal{L}\\
\left(i,j,k,l,m\right)\in Q_{p,q}\\
v\in\left\{ 0,1\right\} }
}l\pi_{i,j,k,l,m,v} & = & 2\beta_{1}c\enskip;\label{eq:l}\\
\sum_{\substack{\left(p,q\right)\in\mathcal{L}\\
\left(i,j,k,l,m\right)\in Q_{p,q}\\
v\in\left\{ 0,1\right\} }
}m\pi_{i,j,k,l,m,v} & = & \beta_{2}c\enskip.\label{eq:m}\end{eqnarray}

\item all occurrences of light variables are allocated to the 5 regions:\\
$\prod_{\substack{\left(p,q\right)\in\mathcal{L}\\
\left(i,j,k,l,m\right)\in Q_{p,q}\\
v\in\left\{ 0,1\right\} }
}\left({i+j+k \choose i,j,k}{l+m \choose l,m}\right)^{\pi_{i,j,k,l,m,v}n}$ allocations are possible;
\item all occurrences of heavy variables are allocated to the 3 satisfied
regions, which yields ${Hn \choose H_{1}n,H_{2}n,H_{3}n}$ possible
allocations; and we must add the following constraint:\begin{eqnarray}
H_{1}+H_{2}+H_{3} & = & H\enskip.\label{eq:ht}\end{eqnarray}

\item all permutations of occurrences of literals are possible inside the
5 regions:\\
their number is $\left(\beta_{1}cn\right)!\left(2\beta_{2}cn\right)!\left(3\beta_{3}cn\right)!\left(2\beta_{1}cn\right)!\left(\beta_{2}cn\right)!$;
\end{enumerate}
\item total number of configurations: the occurrences of literals can be
in any order: $\left(3cn\right)!$ permutations are possible.
\end{enumerate}
\end{enumerate}
We denote by $\mathcal{P}$ the set of all families $\boldsymbol{\zeta}$
of non negative numbers \begin{equation}
\left(\left(\pi_{i,j,k,l,m,v}\right)_{\substack{\left(p,q\right)\in\mathcal{L}\\
\left(i,j,k,l,m\right)\in Q_{p,q}\\
v\in\left\{ 0,1\right\} }
},\left(H_{1},H_{2},H_{3}\right),\left(\beta_{1},\beta_{2},\beta_{3}\right)\right)\label{eq:polytope}\end{equation}
 satisfying the above constraints; note that $\mathcal{P}$ is convex
(by linearity of constraints). We denote by $\mathcal{I}\left(n\right)$
the intersection of $\mathcal{P}$ with the multiples of $\frac{1}{n}$;
we get the following expression of the first moment: $\mathrm{E}X=\sum_{\boldsymbol{\zeta}\in\mathcal{I}\left(n\right)}T\left(n\right)$
where

\begin{eqnarray}
T\left(n\right) & = & 2^{\tau n}{Hn \choose H_{1}n,H_{2}n,H_{3}n}{cn \choose \beta_{1}cn,\beta_{2}cn,\beta_{3}cn}3^{\left(\beta_{1}+\beta_{2}\right)cn}\nonumber \\
 &  & \cdot\frac{\left(\beta_{1}cn\right)!\left(2\beta_{2}cn\right)!\left(3\beta_{3}cn\right)!\left(2\beta_{1}cn\right)!\left(\beta_{2}cn\right)!}{\left(3cn\right)!}\nonumber \\
 &  & \cdot\prod_{\left(p,q\right)\in\mathcal{L}}{d_{p,q}n \choose \delta_{p,q}n}\prod_{\left(p,q\right)\in\mathcal{L}}{\delta_{p,q}n \choose \dots\pi_{i,j,k,l,m,1}n\dots}_{\substack{\left(i,j,k,l,m\right)\in Q_{p,q}}
}\nonumber \\
 &  & \cdot\prod_{\left(p,q\right)\in\mathcal{L}}{\left(d_{p,q}-\delta_{p,q}\right)n \choose \dots\pi_{i,j,k,l,m,0}n\dots}_{\substack{\left(i,j,k,l,m\right)\in Q_{q,p}}
}\nonumber \\
 &  & \cdot\prod_{\substack{\left(p,q\right)\in\mathcal{L}\\
\left(i,j,k,l,m\right)\in Q_{p,q}\\
v\in\left\{ 0,1\right\} }
}\left(\omega_{i,j,k,l,m,v}{i+j+k \choose i,j,k}{l+m \choose l,m}\right)^{\pi_{i,j,k,l,m,v}n}\enskip.\end{eqnarray}

We get rid of all factorials thanks to the following Stirling's inequalities
due to Batir \cite{Batir2008}: $\left(\frac{k}{e}\right)^{k}\sqrt{2\pi\left(k+\frac{1}{6}\right)}<k!<\left(\frac{k}{e}\right)^{k}\sqrt{2\pi\left(k+\left(\frac{e^{2}}{2\pi}-1\right)\right)}$
.

The boundedness of the set $\mathcal{L}$ of light variables (and
thus the boundedness of the sets $Q_{p,q}$) allows to write that
$T\left(n\right)\leq\mathrm{poly}_{1}\left(n\right)F^{n}$ where

\begin{eqnarray}
F & = & 2^{\tau}\frac{H^{H}}{H_{1}^{H_{1}}H_{2}^{H_{2}}H_{3}^{H_{3}}}\left(\frac{1}{3}\left(2\beta_{1}\right)^{\beta_{1}}\left(2\beta_{2}\right)^{\beta_{2}}\left(3\beta_{3}\right)^{\beta_{3}}\right)^{2c}\nonumber \\
 &  & \prod_{\left(p,q\right)\in\mathcal{L}}d_{p,q}^{d_{p,q}}\prod_{\substack{\left(p,q\right)\in\mathcal{L}\\
\left(i,j,k,l,m\right)\in Q_{p,q}\\
v\in\left\{ 0,1\right\} }
}\left(\omega_{i,j,k,l,m,v}\frac{{i+j+k \choose i,j,k}{l+m \choose l,m}}{\pi_{i,j,k,l,m,v}}\right)^{\pi_{i,j,k,l,m,v}}\enskip.\label{eq:F}\end{eqnarray}

Once again, by the lightness property, $\mathcal{I}\left(n\right)$
consists of a bounded number of variables, each of which can take
at most $n+1$ values (as a multiple of $\frac{1}{n}$ ranging between
$0$ and $1$). It follows that the size of $\mathcal{I}\left(n\right)$
is bounded by a polynomial $\mathrm{poly}_{2}\left(n\right)$. And
since $\mathcal{I}\left(n\right)\subseteq\mathcal{P}$, we have $\mathrm{E}X\leq\mathrm{poly}_{2}\left(n\right)\mathrm{poly}_{1}\left(n\right)\left(\max_{\boldsymbol{\zeta}\in\mathcal{P}}F\right)^{n}$.

\subsection{Maximization of $\ln F$ \label{sub:Maximization}}

This is the technical part of our work. We mainly use the same techniques
as \cite{Diaz2009}.
\begin{enumerate}
\item In order to maximize $\ln F$ under our constraints, we use the standard
Lagrange multipliers technique. This is appendix \ref{sec:General-Lagrange}.
The following equations come from the Lagrange derivations and are
important for our study:\begin{eqnarray}
\pi_{i,j,k,l,m,1} & = & \omega_{i,j,k,l,m,1}{i+j+k \choose i,j,k}{l+m \choose l,m}r_{i+j+k,l+m}x_{1}^{2i}x_{2}^{j}y_{1}^{l}y_{2}^{2m}\enskip\label{eq:pi1}\\
\pi_{i,j,k,l,m,0} & = & \omega_{i,j,k,l,m,0}{i+j+k \choose i,j,k}{l+m \choose l,m}r_{l+m,i+j+k}x_{1}^{2i}x_{2}^{j}y_{1}^{l}y_{2}^{2m}\enskip\label{eq:pi0}\end{eqnarray}

$x_{1}$, $x_{2}$,$y_{1}$ and $y_{2}$ are Lagrange multipliers,
that is positive numbers; moreover $r_{p,q}$ is defined as follows:\begin{eqnarray}
r_{p,q} & = & \frac{d_{p,q}}{A_{p,q}}\enskip;\label{eq:r}\\
A_{p,q} & = & \sum_{\substack{\left(i,j,k,l,m\right)\in Q_{p,q}}
}\omega_{i,j,k,l,m,1}{p \choose i,j,k}{q \choose l,m}x_{1}^{2i}x_{2}^{j}y_{1}^{l}y_{2}^{2m}\nonumber \\
 &  & +\sum_{\substack{\left(i,j,k,l,m\right)\in Q_{q,p}}
}\omega_{i,j,k,l,m,0}{q \choose i,j,k}{p \choose l,m}x_{1}^{2i}x_{2}^{j}y_{1}^{l}y_{2}^{2m}\enskip.\label{eq:Apq}\end{eqnarray}

\item In order to justify the use of this technique we must show that the
function $\ln F$ does not maximize on the boundary of the polytope
of constraints; to do so we show that starting at a boundary point
there is always a {}``good'' direction inside the polytope which
makes $\ln F$ greater. This is appendix \ref{sec:Inspection-boundary}.
\item Finally we must ensure that the solution we found by the Lagrange
multiplier technique is indeed a global maximum; to do so we make
a sweep over different values of the parameters $\beta_{t}$; indeed
when these $\beta_{t}$ are fixed the function $\ln F$ is strictly
concave relative to the remaining variables, thus easier to maximize.
This is appendix \ref{sec:Inspection-interior}.
\end{enumerate}

\subsection{Minimization of Global Weight\label{sub:weights-tuning}}

Let us see how one can minimize $F$ (or equivalently $\ln F$) by
a good choice of the weights. The following reasoning is not rigorous;
we only aim at giving some hints to explain the choice of the weights
we made in section \ref{sec:Selection-of-Solutions}.

Remember that $F$ is given by equation \ref{eq:F}. We want to minimize
$\ln F$ by tuning the weights $\omega_{0,j,k,l,m,v}$, so we are
going to differentiate $\ln F$ with respect to an individual $\omega_{0,j,k,l,m,1}$.
Of course due to the constraints every variable depend on $\omega_{0,j,k,l,m,1}$
in the process of maximizing $\ln F$ under these constraints. But
we consider that the variations on all variables are negligible except
for $\pi_{0,j,k,l,m,1}$ (because of equation \ref{eq:pi1}) and $\pi_{0,l,m,j,k,0}$
(because of equations \ref{eq:pi0} and \ref{eq:neighbors}), so we
can write:\begin{eqnarray}
\frac{\partial\left(\ln F\right)}{\partial\omega_{0,j,k,l,m,1}} & \simeq & \frac{\partial\left(\ln F\right)}{\partial\pi_{0,j,k,l,m,1}}\frac{\partial\pi_{0,j,k,l,m,1}}{\partial\omega_{0,j,k,l,m,1}}+\frac{\partial\left(\ln F\right)}{\partial\pi_{0,l,m,j,k,0}}\frac{\partial\pi_{0,l,m,j,k,0}}{\partial\omega_{0,j,k,l,m,1}}\enskip.\end{eqnarray}

Using equations \ref{eq:pi1}, \ref{eq:pi0} and \ref{eq:neighbors}
we find that:

\begin{eqnarray}
\frac{\partial\left(\ln F\right)}{\partial\omega_{0,j,k,l,m,1}} & \simeq & -{j+k \choose j,k}{l+m \choose l,m}r_{j+k,l+m}x_{2}^{j}y_{1}^{l}y_{2}^{2m}\ln\left(r_{j+k,l+m}x_{2}^{j}y_{1}^{l}y_{2}^{2m}\right)\nonumber \\
 &  & +{j+k \choose j,k}{l+m \choose l,m}r_{j+k,l+m}x_{2}^{l}y_{1}^{j}y_{2}^{2k}\ln\left(r_{j+k,l+m}x_{2}^{l}y_{1}^{j}y_{2}^{2k}\right)\enskip.\end{eqnarray}

Now due to equations \ref{eq:r} and \ref{eq:Apq} and numerical experiments
we make the following approximations:\\
$r_{j+k,l+m}x_{2}^{j}y_{1}^{l}y_{2}^{2m}\ll1$ and $r_{j+k,l+m}x_{2}^{l}y_{1}^{j}y_{2}^{2k}\ll1$.
As the function $x\mapsto x\ln\left(ax\right)$ is strictly decreasing
between $0$ and $\frac{1}{ea}$, we can infer the following property:
$\frac{\partial\left(\ln F\right)}{\partial\omega_{0,j,k,l,m,1}}>0$
iff $x_{2}^{l}y_{1}^{j}y_{2}^{2k}<x_{2}^{j}y_{1}^{l}y_{2}^{2m}$,
i.e. $\left(\frac{y_{1}}{x_{2}}\right)^{j-l}\left(y_{2}^{2}\right)^{k-m}<1$.

Now let us consider we are at the minimum point of $\ln F$. If $\frac{\partial\ln\left(F\right)}{\partial\omega_{0,j,k,l,m,1}}\neq0$,
then $\omega_{0,j,k,l,m,1}$ must be at the boundary, i.e. $0$ or
$1$. 

$\frac{\partial\left(\ln F\right)}{\partial\omega_{0,j,k,l,m,1}}>0$
iff $\alpha_{1}\rho_{j,l}+\alpha_{3}\rho_{k,m}<0$, where $\alpha_{1}=\ln\frac{y_{1}}{x_{2}}$
and $\alpha_{3}=\ln\left(y_{2}^{2}\right)$. Thus:
\begin{enumerate}
\item if $\alpha_{1}\rho_{j,l}+\alpha_{3}\rho_{k,m}<0$, then $\omega_{0,j,k,l,m,1}=0$; 
\item if $\alpha_{1}\rho_{j,l}+\alpha_{3}\rho_{k,m}>0$, then $\omega_{0,j,k,l,m,1}=1$;
\item if $\alpha_{1}\rho_{j,l}+\alpha_{3}\rho_{k,m}=0$, nothing can be
said about $\omega_{0,j,k,l,m,1}$.
\end{enumerate}
What about $\omega_{0,j,k,l,m,0}$?
\begin{enumerate}
\item if $\alpha_{1}\rho_{j,l}+\alpha_{3}\rho_{k,m}<0$, then $\alpha_{1}\rho_{l,j}+\alpha_{3}\rho_{m,k}>0$,
thus $\omega_{0,l,m,j,k,1}=1$, so $\omega_{0,j,k,l,m,0}=0$;
\item if $\alpha_{1}\rho_{j,l}+\alpha_{3}\rho_{k,m}>0$, then by the same
argument, $\omega_{0,j,k,l,m,0}=1$;
\item if $\alpha_{1}\rho_{j,l}+\alpha_{3}\rho_{k,m}=0$, nothing can be
said about $\omega_{0,j,k,l,m,0}$.
\end{enumerate}

\section{Conclusion}

We hope that the new track we opened will help gain some more insight
and some more decimals in the quest of the \threesat{} threshold.
In particular note that we required the relation $>$ between solutions
to be circuit-free although this might not be necessary; indeed we
only used the fact that this relation had at least one minimal element.
The same remark holds for the constraints we put onto the weights
of two neighboring solutions as introduced in equation \ref{eq:neighbors},
since this might be too strong. Thus there may be better orientations
or weighting schemes than ours.

\bibliographystyle{splncs}
\bibliography{FM-dpq-ijklm-weights-SAT2010}

\begin{thebibliography}{10}

\bibitem{Mitchell1992}
Mitchell, D., Selman, B., Levesque, H.:
\newblock {Hard and easy distributions of SAT problems}.
\newblock In: Proceedings of the National Conference on Artificial
  Intelligence, AAAI (1992)  459–459

\bibitem{Boufkhad2005}
Boufkhad, Y., Dubois, O., Interian, Y., Selman, B.:
\newblock {Regular Random k-\{SAT\}: Properties of Balanced Formulas}.
\newblock J. Autom. Reasoning \textbf{35}(1-3) (2005)  181--200

\bibitem{Friedgut1999}
Friedgut, E., Bourgain, J.:
\newblock {Sharp thresholds of graph properties, and the k-sat problem}.
\newblock Journal of the American Mathematical Society \textbf{12}(4) (1999)
  1017–1054

\bibitem{Achlioptas2003}
Achlioptas, D., Peres, Y.:
\newblock {The Threshold for Random k-SAT is 2\^{}k ln2 - O(k)}.
\newblock JAMS: Journal of the American Mathematical Society \textbf{17} (2004)
   947----973

\bibitem{ProbabilisticanalysisoftheDavisPutnamprocedureforsolvingthesatisfiabi%
lityproblem::1983::FrancoPaull}
Franco, J., Paull, M.:
\newblock Probabilistic analysis of the davis putnam procedure for solving the
  satisfiability problem.
\newblock Discrete Appl.~Math. \textbf{5} (1983)  77--87

\bibitem{OnRandom3-sat.::1995::MaftouhiVega}
Maftouhi, A.E., de~la Vega, W.F.:
\newblock On random 3-sat.
\newblock Combinatorics, Probability {\&} Computing \textbf{4} (1995)  189--195

\bibitem{Kamath1995}
Kamath, A., Motwani, R., Palem, K., Spirakis, P.:
\newblock {Tail bounds for occupancy and the satisfiability threshold
  conjecture}.
\newblock Random Structures and Algorithms \textbf{7}(1) (1995)  59–80

\bibitem{Dubois1997}
Dubois, O., Boufkhad, Y.:
\newblock {A General Upper Bound for the Satisfiability Threshold of Random
  r-\{SAT\} Formulae.}
\newblock J. Algorithms \textbf{24}(2) (1997)  395--420

\bibitem{Kirousis1998}
Kirousis, L.M., Kranakis, E., Krizanc, D., Stamatiou, Y.C.:
\newblock {Approximating the unsatisfiability threshold of random formulas}.
\newblock Random Structures and Algorithms \textbf{12}(3) (1998)  253--269

\bibitem{Dubois2000}
Dubois, O., Boufkhad, Y., Mandler, J.:
\newblock {Typical random 3-SAT formulae and the satisfiability threshold}.
\newblock In: Proceedings of the eleventh annual ACM-SIAM symposium on Discrete
  algorithms, Society for Industrial and Applied Mathematics (2000)  126----127

\bibitem{Diaz2009}
D\'{\i}az, J., Kirousis, L., Mitsche, D., P\'{e}rez-Gim\'{e}nez, X.:
\newblock {On the satisfiability threshold of formulas with three literals per
  clause}.
\newblock Theoretical Computer Science \textbf{410}(30-32) (2009)  2920--2934

\bibitem{Dubois2003}
Dubois, O., Boufkhad, Y., Mandler, J.:
\newblock {Typical random 3-SAT formulae and the satisfiability threshold}.
\newblock Technical report, ECCC (2003)

\bibitem{Ardila2009a}
Ardila, F., Maneva, E.N.:
\newblock {Pruning processes and a new characterization of convex geometries}.
\newblock Discrete Mathematics \textbf{309}(10) (2009)  3083--3091

\bibitem{Batir2008}
Batir, N.:
\newblock {Inequalities for the gamma function}.
\newblock Archiv der Mathematik \textbf{91}(6) (2008)  554--563

\end{thebibliography}

\appendix

\section*{Appendices}

Let us recall that we want to maximize the function:

\begin{eqnarray}
F & = & 2^{\tau}\frac{H^{H}}{H_{1}^{H_{1}}H_{2}^{H_{2}}H_{3}^{H_{3}}}\left(\frac{1}{3}\left(2\beta_{1}\right)^{\beta_{1}}\left(2\beta_{2}\right)^{\beta_{2}}\left(3\beta_{3}\right)^{\beta_{3}}\right)^{2c}\nonumber \\
 &  & \prod_{\left(p,q\right)\in\mathcal{L}}d_{p,q}^{d_{p,q}}\prod_{\substack{\left(p,q\right)\in\mathcal{L}\\
\left(i,j,k,l,m\right)\in Q_{p,q}\\
v\in\left\{ 0,1\right\} }
}\left(\omega_{i,j,k,l,m,v}\frac{{i+j+k \choose i,j,k}{l+m \choose l,m}}{\pi_{i,j,k,l,m,v}}\right)^{\pi_{i,j,k,l,m,v}}\enskip.\end{eqnarray}

on variables $\left(\pi_{i,j,k,l,m,v}\right)_{\substack{\left(p,q\right)\in\mathcal{L}\\
\left(i,j,k,l,m\right)\in Q_{p,q}\\
v\in\left\{ 0,1\right\} }
},\left(H_{1},H_{2},H_{3}\right),\left(\beta_{1},\beta_{2},\beta_{3}\right)$ subject to the following constraints:\begin{eqnarray}
\beta_{1}+\beta_{2}+\beta_{3} & = & 1\label{eq:bt}\\
\sum_{\substack{\left(i,j,k,l,m\right)\in Q_{p,q}}
}\pi_{i,j,k,l,m,1}+\sum_{\substack{\left(i,j,k,l,m\right)\in Q_{q,p}}
}\pi_{i,j,k,l,m,0} & = & d_{p,q}\label{eq:dpq}\\
H_{1}+H_{2}+H_{3} & = & \sum_{\left(p,q\right)\in\mathcal{H}}\left(p+q\right)d_{p,q}\label{eq:ht}\\
\sum_{\substack{\left(p,q\right)\in\mathcal{L}\\
\left(i,j,k,l,m\right)\in Q_{p,q}\\
v\in\left\{ 0,1\right\} }
}i\pi_{i,j,k,l,m,v}+H_{1} & = & \beta_{1}c\label{eq:i}\\
\sum_{\substack{\left(p,q\right)\in\mathcal{L}\\
\left(i,j,k,l,m\right)\in Q_{p,q}\\
v\in\left\{ 0,1\right\} }
}j\pi_{i,j,k,l,m,v}+H_{2} & = & 2\beta_{2}c\label{eq:j}\\
\sum_{\substack{\left(p,q\right)\in\mathcal{L}\\
\left(i,j,k,l,m\right)\in Q_{p,q}\\
v\in\left\{ 0,1\right\} }
}k\pi_{i,j,k,l,m,v}+H_{3} & = & 3\beta_{3}c\label{eq:k}\\
\sum_{\substack{\left(p,q\right)\in\mathcal{L}\\
\left(i,j,k,l,m\right)\in Q_{p,q}\\
v\in\left\{ 0,1\right\} }
}l\pi_{i,j,k,l,m,v} & = & 2\beta_{1}c\label{eq:l}\\
\sum_{\substack{\left(p,q\right)\in\mathcal{L}\\
\left(i,j,k,l,m\right)\in Q_{p,q}\\
v\in\left\{ 0,1\right\} }
}m\pi_{i,j,k,l,m,v} & = & \beta_{2}c\label{eq:m}\end{eqnarray}

To perform such a maximization we use the standard technique of Lagrange
multipliers.

\section{Resolution of the Global Lagrange Multipliers Problem\label{sec:General-Lagrange}}

\subsection{Elimination of redundant constraints}

The first thing to do is to remove redundant constraints. It appears
that \eg{} constraint \eqref{eq:k} is redundant with constraints
\eqref{eq:i}, \eqref{eq:k}, \eqref{eq:l}, \eqref{eq:m}, because
summing these 5 equations and using the previous ones \eqref{eq:bt},
\eqref{eq:ht}, \eqref{eq:dpq} gives a tautology:\begin{eqnarray*}
\sum_{\substack{\left(p,q\right)\in\mathcal{L}\\
\left(i,j,k,l,m\right)\in Q_{p,q}\\
v\in\left\{ 0,1\right\} }
}\left(i+j+k+l+m\right)\pi_{i,j,k,l,m,v}+H_{1}+H_{2}+H_{3} & = & 3\left(\beta_{1}+\beta_{2}+\beta_{3}\right)c\end{eqnarray*}
\begin{eqnarray*}
\sum_{\substack{\left(p,q\right)\in\mathcal{L}\\
\left(i,j,k,l,m\right)\in Q_{p,q}}
}\left(\left(i+j+k\right)\pi_{i,j,k,l,m,1}+\left(l+m\right)\pi_{i,j,k,l,m,1}+\left(i+j+k\right)\pi_{i,j,k,l,m,0}+\left(l+m\right)\pi_{i,j,k,l,m,0}\right)+H & = & 3c\end{eqnarray*}
\begin{eqnarray*}
\sum_{\substack{\left(p,q\right)\in\mathcal{L}\\
\left(i,j,k,l,m\right)\in Q_{p,q}}
}\left(\left(i+j+k\right)\pi_{i,j,k,l,m,1}+\left(l+m\right)\pi_{i,j,k,l,m,1}\right)\\
+\sum_{\substack{\left(p,q\right)\in\mathcal{L}\\
\left(i,j,k,l,m\right)\in Q_{q,p}}
}\left(\left(i+j+k\right)\pi_{i,j,k,l,m,0}+\left(l+m\right)\pi_{i,j,k,l,m,0}\right)+H & = & 3c\end{eqnarray*}
\begin{eqnarray*}
\sum_{\substack{\left(p,q\right)\in\mathcal{L}\\
\left(i,j,k,l,m\right)\in Q_{p,q}}
}\left(p\pi_{i,j,k,l,m,1}+q\pi_{i,j,k,l,m,1}\right)+\sum_{\substack{\left(p,q\right)\in\mathcal{L}\\
\left(i,j,k,l,m\right)\in Q_{q,p}}
}\left(q\pi_{i,j,k,l,m,0}+p\pi_{i,j,k,l,m,0}\right)+H & = & 3c\end{eqnarray*}

\begin{eqnarray*}
\sum_{\left(p,q\right)\in\mathcal{L}}\left(p+q\right)\left(\sum_{\substack{\left(i,j,k,l,m\right)\in Q_{p,q}}
}\pi_{i,j,k,l,m,1}+\sum_{\substack{\left(i,j,k,l,m\right)\in Q_{q,p}}
}\pi_{i,j,k,l,m,0}\right)+H & = & 3c\end{eqnarray*}
\begin{eqnarray*}
\sum_{\left(p,q\right)\in\mathcal{L}}\left(p+q\right)d_{p,q} & +\sum_{\left(p,q\right)\in\mathcal{H}}\left(p+q\right)d_{p,q}= & 3c\end{eqnarray*}

which was a requirement we made on $\left(d_{p,q}\right)$.

Thus we get rid of constraint \eqref{eq:k} and there remain 7 constraints.

\subsection{Definition of the Lagrangian}

\begin{eqnarray*}
\Lambda & = & \tau\ln2+H\ln H-H_{1}\ln\left(\frac{H_{1}}{e}\right)-H_{2}\ln\left(\frac{H_{2}}{e}\right)-H_{3}\ln\left(\frac{H_{3}}{e}\right)-H\\
 &  & +\sum_{\left(p,q\right)\in\mathcal{L}}d_{p,q}\ln d_{p,q}+\sum_{\substack{\left(p,q\right)\in\mathcal{L}\\
\left(i,j,k,l,m\right)\in Q_{p,q}\\
v\in\left\{ 0,1\right\} }
}\pi_{i,j,k,l,m,v}\ln\left(\omega_{i,j,k,l,m,v}\frac{e{i+j+k \choose i,j,k}{l+m \choose l,m}}{\pi_{i,j,k,l,m,v}}\right)-1\\
 &  & -2c\ln3+2c\beta_{1}\ln\left(2\frac{\beta_{1}}{e}\right)+2c\beta_{2}\ln\left(2\frac{\beta_{2}}{e}\right)+2c\beta_{3}\ln\left(3\frac{\beta_{3}}{e}\right)+2c\\
 &  & +\left(2c\ln b\right)\left(\beta_{1}+\beta_{2}+\beta_{3}-1\right)\\
 &  & +\sum_{\left(p,q\right)\in\mathcal{L}}\left(\ln r_{p,q}\right)\left(\sum_{\substack{\left(i,j,k,l,m\right)\in Q_{p,q}}
}\pi_{i,j,k,l,m,1}+\sum_{\substack{\left(i,j,k,l,m\right)\in Q_{q,p}}
}\pi_{i,j,k,l,m,0}-d_{p,q}\right)\\
 &  & +\left(\ln h\right)\left(H_{1}+H_{2}+H_{3}-\sum_{\left(p,q\right)\in\mathcal{H}}\left(p+q\right)d_{p,q}\right)\\
 &  & +\left(2\ln x_{1}\right)\left(\sum_{\substack{\left(p,q\right)\in\mathcal{L}\\
\left(i,j,k,l,m\right)\in Q_{p,q}\\
v\in\left\{ 0,1\right\} }
}i\pi_{i,j,k,l,m,v}+H_{1}-\beta_{1}c\right)+\left(\ln x_{2}\right)\left(\sum_{\substack{\left(p,q\right)\in\mathcal{L}\\
\left(i,j,k,l,m\right)\in Q_{p,q}\\
v\in\left\{ 0,1\right\} }
}j\pi_{i,j,k,l,m,v}+H_{2}-2\beta_{2}c\right)\\
 &  & +\left(\ln y_{1}\right)\left(\sum_{\substack{\left(p,q\right)\in\mathcal{L}\\
\left(i,j,k,l,m\right)\in Q_{p,q}\\
v\in\left\{ 0,1\right\} }
}l\pi_{i,j,k,l,m,v}-2\beta_{1}c\right)+\left(2\ln y_{2}\right)\left(\sum_{\substack{\left(p,q\right)\in\mathcal{L}\\
\left(i,j,k,l,m\right)\in Q_{p,q}\\
v\in\left\{ 0,1\right\} }
}m\pi_{i,j,k,l,m,v}-\beta_{2}c\right)\end{eqnarray*}

\subsection{Derivatives with Respect to $\pi_{i,j,k,l,m,v}$}

\begin{eqnarray*}
\frac{\partial\Lambda}{\partial\pi_{i,j,k,l,m,1}} & = & \ln\omega_{i,j,k,l,m,1}+\ln\left({i+j+k \choose i,j,k}{l+m \choose l,m}\right)-\ln\pi_{i,j,k,l,m,v}+\ln r_{i+j+k,l+m}\\
 &  & +2i\ln x_{1}+j\ln x_{2}+l\ln y_{1}+2m\ln y_{2}\\
\frac{\partial\Lambda}{\partial\pi_{i,j,k,l,m,0}} & = & \ln\omega_{i,j,k,l,m,0}+\ln\left({i+j+k \choose i,j,k}{l+m \choose l,m}\right)-\ln\pi_{i,j,k,l,m,v}+\ln r_{l+m,i+j+k}\\
 &  & +2i\ln x_{1}+j\ln x_{2}+l\ln y_{1}+2m\ln y_{2}\end{eqnarray*}

\begin{eqnarray*}
\pi_{i,j,k,l,m,1} & = & \omega_{i,j,k,l,m,1}{i+j+k \choose i,j,k}{l+m \choose l,m}r_{i+j+k,l+m}x_{1}^{2i}x_{2}^{j}y_{1}^{l}y_{2}^{2m}\\
\pi_{i,j,k,l,m,0} & = & \omega_{i,j,k,l,m,0}{i+j+k \choose i,j,k}{l+m \choose l,m}r_{l+m,i+j+k}x_{1}^{2i}x_{2}^{j}y_{1}^{l}y_{2}^{2m}\end{eqnarray*}

The $r_{p,q}$ contraints become:\begin{eqnarray*}
\sum_{\substack{\left(i,j,k,l,m\right)\in Q_{p,q}}
}\omega_{i,j,k,l,m,1}{i+j+k \choose i,j,k}{l+m \choose l,m}r_{i+j+k,l+m}x_{1}^{2i}x_{2}^{j}y_{1}^{l}y_{2}^{2m}\\
+\sum_{\substack{\left(i,j,k,l,m\right)\in Q_{q,p}}
}\omega_{i,j,k,l,m,0}{i+j+k \choose i,j,k}{l+m \choose l,m}r_{l+m,i+j+k}x_{1}^{2i}x_{2}^{j}y_{1}^{l}y_{2}^{2m} & = & d_{p,q}\end{eqnarray*}

Let us introduce \[
A_{p,q}=\sum_{\substack{\left(i,j,k,l,m\right)\in Q_{p,q}}
}\omega_{i,j,k,l,m,1}{p \choose i,j,k}{q \choose l,m}x_{1}^{2i}x_{2}^{j}y_{1}^{l}y_{2}^{2m}+\sum_{\substack{\left(i,j,k,l,m\right)\in Q_{q,p}}
}\omega_{i,j,k,l,m,0}{q \choose i,j,k}{p \choose l,m}x_{1}^{2i}x_{2}^{j}y_{1}^{l}y_{2}^{2m}\]

We have:\begin{eqnarray*}
r_{p,q}A_{p,q} & = & d_{p,q}\end{eqnarray*}
\begin{eqnarray*}
r_{p,q} & = & \frac{d_{p,q}}{A_{p,q}}\end{eqnarray*}

Thus\begin{eqnarray*}
\pi_{i,j,k,l,m,1} & = & \omega_{i,j,k,l,m,1}{i+j+k \choose i,j,k}{l+m \choose l,m}\frac{d_{i+j+k,l+m}}{A_{i+j+k,l+m}}x_{1}^{2i}x_{2}^{j}y_{1}^{l}y_{2}^{2m}\\
\pi_{i,j,k,l,m,0} & = & \omega_{i,j,k,l,m,0}{i+j+k \choose i,j,k}{l+m \choose l,m}\frac{d_{l+m,i+j+k}}{A_{l+m,i+j+k}}x_{1}^{2i}x_{2}^{j}y_{1}^{l}y_{2}^{2m}\end{eqnarray*}

\subsection{Derivatives with Respect to $\beta_{t}$}

\begin{eqnarray*}
\frac{\partial\Lambda}{\partial\beta_{1}} & = & 2c\ln2+2c\ln\beta_{1}+2c\ln b-2c\ln x_{1}-2c\ln y_{1}\\
\frac{\partial\Lambda}{\partial\beta_{2}} & = & 2c\ln2+2c\ln\beta_{2}+2c\ln b-2c\ln x_{2}-2c\ln y_{2}\\
\frac{\partial\Lambda}{\partial\beta_{3}} & = & 2c\ln3+2c\ln\beta_{3}+2c\ln b\end{eqnarray*}

\begin{eqnarray*}
\beta_{1} & = & \frac{x_{1}y_{1}}{2b}\\
\beta_{2} & = & \frac{x_{2}y_{2}}{2b}\\
\beta_{3} & = & \frac{1}{3b}\end{eqnarray*}

The $b$ constraint becomes:

\begin{eqnarray*}
b & = & \frac{x_{1}y_{1}}{2}+\frac{x_{2}y_{2}}{2}+\frac{1}{3}\end{eqnarray*}

\subsection{Derivatives with Respect to $H_{t}$}

\begin{eqnarray*}
\frac{\partial\Lambda}{\partial H_{1}} & = & -\ln H_{1}+\ln h+2\ln x_{1}\\
\frac{\partial\Lambda}{\partial H_{2}} & = & -\ln H_{2}+\ln h+\ln x_{2}\\
\frac{\partial\Lambda}{\partial H_{3}} & = & -\ln H_{3}+\ln h\end{eqnarray*}

\begin{eqnarray*}
H_{1} & = & hx_{1}^{2}\\
H_{2} & = & hx_{2}\\
H_{3} & = & h\end{eqnarray*}

The $h$ constraint becomes then:

\begin{eqnarray*}
h\left(x_{1}^{2}+x_{2}+1\right) & = & H\end{eqnarray*}

\begin{eqnarray*}
h & = & \frac{H}{x_{1}^{2}+x_{2}+1}\end{eqnarray*}

Thus:\begin{eqnarray*}
H_{1} & = & \frac{Hx_{1}^{2}}{x_{1}^{2}+x_{2}+1}\\
H_{2} & = & \frac{Hx_{2}}{x_{1}^{2}+x_{2}+1}\\
H_{3} & = & \frac{H}{x_{1}^{2}+x_{2}+1}\end{eqnarray*}

\subsection{Further Simplifications}
\begin{itemize}
\item 1st Moment:

\begin{eqnarray*}
F & = & 2^{\tau}\frac{H^{H}}{H_{1}^{H_{1}}H_{2}^{H_{2}}H_{3}^{H_{3}}}\prod_{\left(p,q\right)\in\mathcal{L}}d_{p,q}^{d_{p,q}}\prod_{\substack{\left(p,q\right)\in\mathcal{L}\\
\left(i,j,k,l,m\right)\in Q_{p,q}\\
v\in\left\{ 0,1\right\} }
}\left(\omega_{i,j,k,l,m,v}\frac{{i+j+k \choose i,j,k}{l+m \choose l,m}}{\pi_{i,j,k,l,m,v}}\right)^{\pi_{i,j,k,l,m,v}}\left(\frac{1}{3}\left(2\beta_{1}\right)^{\beta_{1}}\left(2\beta_{2}\right)^{\beta_{2}}\left(3\beta_{3}\right)^{\beta_{3}}\right)^{2c}\end{eqnarray*}

\begin{eqnarray*}
F & = & \prod_{\left(p,q\right)\in\mathcal{L}}d_{p,q}^{d_{p,q}}\prod_{\substack{\left(p,q\right)\in\mathcal{L}\\
\left(i,j,k,l,m\right)\in Q_{p,q}}
}\left(\frac{A_{i+j+k,l+m}}{d_{i+j+k,l+m}x_{1}^{2i}x_{2}^{j}y_{1}^{l}y_{2}^{2m}}\right)^{\pi_{i,j,k,l,m,1}}\prod_{\substack{\left(p,q\right)\in\mathcal{L}\\
\left(i,j,k,l,m\right)\in Q_{q,p}}
}\left(\frac{A_{l+m,i+j+k}}{d_{l+m,i+j+k}x_{1}^{2i}x_{2}^{j}y_{1}^{l}y_{2}^{2m}}\right)^{\pi_{i,j,k,l,m,0}}\\
 &  & \cdot2^{\tau}\frac{H^{H}}{\left(hx_{1}^{2}\right)^{H_{1}}\left(hx_{2}\right)^{H_{2}}\left(h\right)^{H_{3}}}\left(\frac{1}{3b}\left(x_{1}y_{1}\right)^{\beta_{1}}\left(x_{2}y_{2}\right)^{\beta_{2}}\right)^{2c}\end{eqnarray*}
\begin{eqnarray*}
F & = & \prod_{\left(p,q\right)\in\mathcal{L}}\left(d_{p,q}^{d_{p,q}}\prod_{\left(i,j,k,l,m\right)\in Q_{p,q}}\left(\frac{A_{p,q}}{d_{p,q}x_{1}^{2i}x_{2}^{j}y_{1}^{l}y_{2}^{2m}}\right)^{\pi_{i,j,k,l,m,1}}\prod_{\left(i,j,k,l,m\right)\in Q_{q,p}}\left(\frac{A_{p,q}}{d_{p,q}x_{1}^{2i}x_{2}^{j}y_{1}^{l}y_{2}^{2m}}\right)^{\pi_{i,j,k,l,m,0}}\right)\\
 &  & \cdot2^{\tau}H^{H}h^{-\left(H_{1}+H_{2}+H_{3}\right)}x_{1}^{-2H_{1}}x_{2}^{-H_{2}}\left(\frac{1}{3b}x_{1}^{\beta_{1}}y_{1}^{\beta_{1}}x_{2}^{\beta_{2}}y_{2}^{\beta_{2}}\right)^{2c}\end{eqnarray*}
\begin{eqnarray*}
F & = & 2^{\tau}\left(\frac{H}{h}\right)^{H}\prod_{\left(p,q\right)\in\mathcal{L}}\left(d_{p,q}^{d_{p,q}}\left(\frac{A_{p,q}}{d_{p,q}}\right)^{\sum_{\substack{\left(i,j,k,l,m\right)\in Q_{p,q}}
}\pi_{i,j,k,l,m,1}+\sum_{\substack{\left(i,j,k,l,m\right)\in Q_{q,p}}
}\pi_{i,j,k,l,m,0}}\right)\\
 &  & \cdot x_{1}^{-2\left(\sum_{\substack{\left(i,j,k,l,m\right)\in Q_{n}\\
v\in\left\{ 0,1\right\} }
}i\pi_{i,j,k,l,m,v}+H_{1}\right)}x_{2}^{-\left(\sum_{\substack{\left(i,j,k,l,m\right)\in Q_{n}\\
v\in\left\{ 0,1\right\} }
}j\pi_{i,j,k,l,m,v}+H_{2}\right)}\\
 &  & y_{1}^{-\sum_{\substack{\left(i,j,k,l,m\right)\in Q_{n}\\
v\in\left\{ 0,1\right\} }
}l\pi_{i,j,k,l,m,v}}y_{2}^{-2\sum_{\substack{\left(i,j,k,l,m\right)\in Q_{n}\\
v\in\left\{ 0,1\right\} }
}m\pi_{i,j,k,l,m,v}}\\
 &  & \cdot\left(\frac{1}{3b}x_{1}^{\beta_{1}}y_{1}^{\beta_{1}}x_{2}^{\beta_{2}}y_{2}^{\beta_{2}}\right)^{2c}\end{eqnarray*}
\begin{eqnarray*}
F & = & 2^{\tau}\left(\frac{H}{h}\right)^{H}\prod_{\left(p,q\right)\in\mathcal{L}}\left(d_{p,q}^{d_{p,q}}\left(\frac{A_{p,q}}{d_{p,q}}\right)^{d_{p,q}}\right)x_{1}^{-2\beta_{1}c}x_{2}^{-2\beta_{2}c}y_{1}^{-2\beta_{1}c}y_{2}^{-2\beta_{2}c}\left(\frac{1}{3b}x_{1}^{\beta_{1}}y_{1}^{\beta_{1}}x_{2}^{\beta_{2}}y_{2}^{\beta_{2}}\right)^{2c}\end{eqnarray*}
\begin{eqnarray*}
F & = & 2^{\tau}\left(x_{1}^{2}+x_{2}+1\right)^{H}\prod_{\left(p,q\right)\in\mathcal{L}}\left(A_{p,q}^{d_{p,q}}\right)\left(\frac{3x_{1}y_{1}}{2}+\frac{3x_{2}y_{2}}{2}+1\right)^{-2c}\end{eqnarray*}

\item Remaining constraints:

\begin{eqnarray*}
\sum_{\left(p,q\right)\in\mathcal{L}}\frac{d_{p,q}}{A_{p,q}}\sum_{\substack{\left(i,j,k,l,m\right)\in Q_{p,q}}
}i{p \choose i,j,k}{q \choose l,m}x_{1}^{2i}x_{2}^{j}y_{1}^{l}y_{2}^{2m}\omega_{i,j,k,l,m,1}\\
+\sum_{\left(p,q\right)\in\mathcal{L}}\frac{d_{p,q}}{A_{p,q}}\sum_{\substack{\left(i,j,k,l,m\right)\in Q_{q,p}}
}i{q \choose i,j,k}{p \choose l,m}x_{1}^{2i}x_{2}^{j}y_{1}^{l}y_{2}^{2m}\omega_{i,j,k,l,m,0}+\frac{Hx_{1}^{2}}{x_{1}^{2}+x_{2}+1} & = & \frac{x_{1}y_{1}c}{2\left(\frac{x_{1}y_{1}}{2}+\frac{x_{2}y_{2}}{2}+\frac{1}{3}\right)}\\
\sum_{\left(p,q\right)\in\mathcal{L}}\frac{d_{p,q}}{A_{p,q}}\sum_{\substack{\left(i,j,k,l,m\right)\in Q_{p,q}}
}j{p \choose i,j,k}{q \choose l,m}x_{1}^{2i}x_{2}^{j}y_{1}^{l}y_{2}^{2m}\omega_{i,j,k,l,m,1}\\
+\sum_{\left(p,q\right)\in\mathcal{L}}\frac{d_{p,q}}{A_{p,q}}\sum_{\substack{\left(i,j,k,l,m\right)\in Q_{q,p}}
}j{q \choose i,j,k}{p \choose l,m}x_{1}^{2i}x_{2}^{j}y_{1}^{l}y_{2}^{2m}\omega_{i,j,k,l,m,0}+\frac{Hx_{2}}{x_{1}^{2}+x_{2}+1} & = & \frac{x_{2}y_{2}c}{\left(\frac{x_{1}y_{1}}{2}+\frac{x_{2}y_{2}}{2}+\frac{1}{3}\right)}\\
\sum_{\left(p,q\right)\in\mathcal{L}}\frac{d_{p,q}}{A_{p,q}}\sum_{\substack{\left(i,j,k,l,m\right)\in Q_{p,q}}
}l{p \choose i,j,k}{q \choose l,m}x_{1}^{2i}x_{2}^{j}y_{1}^{l}y_{2}^{2m}\omega_{i,j,k,l,m,1}\\
+\sum_{\left(p,q\right)\in\mathcal{L}}\frac{d_{p,q}}{A_{p,q}}\sum_{\substack{\left(i,j,k,l,m\right)\in Q_{q,p}}
}l{q \choose i,j,k}{p \choose l,m}x_{1}^{2i}x_{2}^{j}y_{1}^{l}y_{2}^{2m}\omega_{i,j,k,l,m,0} & = & \frac{x_{1}y_{1}c}{\left(\frac{x_{1}y_{1}}{2}+\frac{x_{2}y_{2}}{2}+\frac{1}{3}\right)}\\
\sum_{\left(p,q\right)\in\mathcal{L}}\frac{d_{p,q}}{A_{p,q}}\sum_{\substack{\left(i,j,k,l,m\right)\in Q_{p,q}}
}m{p \choose i,j,k}{q \choose l,m}x_{1}^{2i}x_{2}^{j}y_{1}^{l}y_{2}^{2m}\omega_{i,j,k,l,m,1}\\
+\sum_{\left(p,q\right)\in\mathcal{L}}\frac{d_{p,q}}{A_{p,q}}\sum_{\substack{\left(i,j,k,l,m\right)\in Q_{q,p}}
}m{q \choose i,j,k}{p \choose l,m}x_{1}^{2i}x_{2}^{j}y_{1}^{l}y_{2}^{2m}\omega_{i,j,k,l,m,0} & = & \frac{x_{2}y_{2}c}{2\left(\frac{x_{1}y_{1}}{2}+\frac{x_{2}y_{2}}{2}+\frac{1}{3}\right)}\end{eqnarray*}
\begin{eqnarray*}
\sum_{\left(p,q\right)\in\mathcal{L}}\frac{d_{p,q}}{A_{p,q}}\frac{x_{1}}{2}\frac{\partial A_{p,q}}{\partial x_{1}}+\frac{Hx_{1}^{2}}{x_{1}^{2}+x_{2}+1} & = & \frac{x_{1}y_{1}c}{2\left(\frac{x_{1}y_{1}}{2}+\frac{x_{2}y_{2}}{2}+\frac{1}{3}\right)}\\
\sum_{\left(p,q\right)\in\mathcal{L}}\frac{d_{p,q}}{A_{p,q}}x_{2}\frac{\partial A_{p,q}}{\partial x_{2}}+\frac{Hx_{2}}{x_{1}^{2}+x_{2}+1} & = & \frac{x_{2}y_{2}c}{\left(\frac{x_{1}y_{1}}{2}+\frac{x_{2}y_{2}}{2}+\frac{1}{3}\right)}\\
\sum_{\left(p,q\right)\in\mathcal{L}}\frac{d_{p,q}}{A_{p,q}}y_{1}\frac{\partial A_{p,q}}{\partial y_{1}} & = & \frac{x_{1}y_{1}c}{\left(\frac{x_{1}y_{1}}{2}+\frac{x_{2}y_{2}}{2}+\frac{1}{3}\right)}\\
\sum_{\left(p,q\right)\in\mathcal{L}}\frac{d_{p,q}}{A_{p,q}}\frac{y_{2}}{2}\frac{\partial A_{p,q}}{\partial y_{2}} & = & \frac{x_{2}y_{2}c}{2\left(\frac{x_{1}y_{1}}{2}+\frac{x_{2}y_{2}}{2}+\frac{1}{3}\right)}\end{eqnarray*}

Then we introduce $Z=\prod_{\left(p,q\right)\in\mathcal{L}}A_{p,q}^{d_{p,q}}$
and $Y=\ln Z$:\begin{eqnarray*}
\frac{\partial Y}{\partial x_{1}}+\frac{2Hx_{1}}{x_{1}^{2}+x_{2}+1} & = & \frac{y_{1}c}{\left(\frac{x_{1}y_{1}}{2}+\frac{x_{2}y_{2}}{2}+\frac{1}{3}\right)}\\
\frac{\partial Y}{\partial x_{2}}+\frac{H}{x_{1}^{2}+x_{2}+1} & = & \frac{y_{2}c}{\left(\frac{x_{1}y_{1}}{2}+\frac{x_{2}y_{2}}{2}+\frac{1}{3}\right)}\\
\frac{\partial Y}{\partial y_{1}} & = & \frac{x_{1}c}{\left(\frac{x_{1}y_{1}}{2}+\frac{x_{2}y_{2}}{2}+\frac{1}{3}\right)}\\
\frac{\partial Y}{\partial y_{2}} & = & \frac{x_{2}c}{\left(\frac{x_{1}y_{1}}{2}+\frac{x_{2}y_{2}}{2}+\frac{1}{3}\right)}\end{eqnarray*}

\end{itemize}
To solve these equations we used \mathematica. The bound we obtained
for $c$ are summed up in table \eqref{tab:general}.%
\begin{table}
\caption{Summary of results obtained by Lagrange's method.\label{tab:general}}

\centering{}\begin{tabular}{|>{\centering}p{0.12\paperwidth}|>{\centering}p{0.1\paperwidth}|>{\centering}p{0.1\paperwidth}|>{\centering}p{0.1\paperwidth}|>{\centering}p{0.13\paperwidth}|}
\hline 
model & standard & balanced signs & balanced occurrences & balanced signs and occurrences\tabularnewline
\hline 
our method

$\left(\alpha\rho_{j,l}+\rho_{k,m},v\right)>\left(0,0\right)$ & $4.500$ & $3.509$ & $4.623$ & $3.546$\tabularnewline
\hline 
our $\alpha$ & $2.00$ & $1.01$ & $2.01$ & $1.01$\tabularnewline
\hline 
$x_{1}$ & $1.0083$ & $1.47787$ & $1.01694$ & $1.57726$\tabularnewline
\hline 
$x_{2}$ & $2.06625$ & $3.09005$ & $2.08256$ & $3.38506$\tabularnewline
\hline 
$y_{1}$ & $2.18256$ & $3.27457$ & $2.19038$ & $3.51076$\tabularnewline
\hline 
$y_{2}$ & $1.01253$ & $1.02742$ & $1.01221$ & $1.045$\tabularnewline
\hline 
$\beta_{1}$ & $0.44373$ & $0.557479$ & $0.445306$ & $0.568436$\tabularnewline
\hline 
$\beta_{2}$ & $0.421847$ & $0.365723$ & $0.421418$ & $0.363128$\tabularnewline
\hline 
$\beta_{3}$ & $0.134422$ & $0.0767974$ & $0.133276$ & $0.0684362$\tabularnewline
\hline
\end{tabular}
\end{table}

\section{Inspection of the Boundary of $\mathcal{P}$\label{sec:Inspection-boundary}}

The boundary of $\mathcal{P}$ is reached when one of the variables
is at $0$. We want to be sure that $F$ cannot be maximized by such
a configuration. Remember that\begin{eqnarray*}
\ln F & = & \tau\ln2+H\ln H-H_{1}\ln\left(\frac{H_{1}}{e}\right)-H_{2}\ln\left(\frac{H_{2}}{e}\right)-H_{3}\ln\left(\frac{H_{3}}{e}\right)-H\\
 &  & +\sum_{\left(p,q\right)\in\mathcal{L}}d_{p,q}\ln d_{p,q}+\sum_{\substack{\left(p,q\right)\in\mathcal{L}\\
\left(i,j,k,l,m\right)\in Q_{p,q}\\
v\in\left\{ 0,1\right\} }
}\pi_{i,j,k,l,m,v}\ln\left(\omega_{i,j,k,l,m,v}\frac{{i+j+k \choose i,j,k}{l+m \choose l,m}}{e\pi_{i,j,k,l,m,v}}\right)+1\\
 &  & -2c\ln3+2c\beta_{1}\ln\left(2\frac{\beta_{1}}{e}\right)+2c\beta_{2}\ln\left(2\frac{\beta_{2}}{e}\right)+2c\beta_{3}\ln\left(3\frac{\beta_{3}}{e}\right)+2c\end{eqnarray*}

If we increase an $H_{t}$ or a $\pi_{i,j,k,l,m,v}$ from $0$ to
a small $\xi>0$ and we change any other non zero variables, then
the variation of $\ln F$ is $f=-\xi\ln\xi+\Theta\left(\xi\right)$
is such that $\frac{f}{\xi}=-\ln\xi+\Theta\left(1\right)\to+\infty$,
so $\ln F$ must increase; but what if we increase a $\beta_{t}$
from $0$ to a $\xi>0$? Then $\frac{f}{\xi}=+\ln\xi+\Theta\left(1\right)\to-\infty$.
Thus the problem at the boundary of $\mathcal{P}$ comes from the
$\beta_{t}$. The technique will be, as in \cite{Dubois2003,Diaz2009},
to make a small move in a well chosen direction in order to circumvent
the negative side-effect of increasing a $\beta_{t}$ which is at
$0$. Such a direction will be referred to as an \emph{increasing
direction}. However we must ensure that such a direction is indeed
in the polytope $\mathcal{P}$. Note that in case we find the direction
by pointing towards another point in $\mathcal{P}$, this property
results from the convexity of $\mathcal{P}$.

We used \mathematica{} to minimize and maximize $\beta_{1}$ under
the above constraints in each model and our corresponding weighting
scheme ; the precise bounds we obtained for $\beta_{1}$ in each model
are summed up in table \eqref{tab:bounds-betas}. Noteworthy is the
fact that $\beta_{1}$ can be neither $0$ nor $1$ (thus we can have
neither $\beta_{1}=0$ nor $\beta_{2}=\beta_{3}=0$).

\begin{table}
\begin{centering}
\caption{Summary of bounds on $\beta_{t}$. \label{tab:bounds-betas}}

\par\end{centering}

\centering{}\begin{tabular}{|>{\centering}p{0.18\columnwidth}|>{\centering}p{0.18\columnwidth}|>{\centering}p{0.18\columnwidth}|>{\centering}p{0.18\columnwidth}|>{\centering}p{0.18\columnwidth}|}
\hline 
model & standard & balanced signs & balanced occurrences & balanced signs and occurrences\tabularnewline
\hline 
our method

$\left(\alpha\rho_{j,l}+\rho_{k,m},v\right)>\left(0,0\right)$ & $4.500$ & $3.509$ & $4.623$ & $3.546$\tabularnewline
\hline 
our $\alpha$ & $2.00$ & $1.01$ & $2.01$ & $1.01$\tabularnewline
\hline 
our bounds for $\beta_{1}$ & $0.177<\beta_{1}<0.912$ & $0.428<\beta_{1}<0.786$ & $0.182<\beta_{1}<0.909$ & $0.5<\beta_{1}<0.75$\tabularnewline
\hline 
(our bounds for $\beta_{2}$ - deductible from above) & $0\leq\beta_{2}<0.823$ & $0\leq\beta_{2}<0.572$ & $0\leq\beta_{2}<0.818$ & $0\leq\beta_{2}<0.5$\tabularnewline
\hline 
our bounds for $\beta_{3}$ & $0\leq\beta_{3}<0.412$ & $0\leq\beta_{3}<0.286$ & $0\leq\beta_{3}<0.409$ & $0\leq\beta_{3}<0.25$\tabularnewline
\hline
\end{tabular}
\end{table}

\begin{enumerate}
\item case where $\beta_{2}=0$: then $H_{2}=0$ and $\pi_{i,j,k,l,m,v}=0$
unless $j=m=0$; we call these variables \emph{forced} as did \cite{Diaz2009};
moreover in the models where there are no heavy variables we consider
variables $H_{t}$ to be forced to $0$ as well.

\begin{itemize}
\item subcase where there is an unforced variable at zero: we find a feasible
point where $\beta_{2}=0$ and all unforced variables are non zero.
Then a move towards this point gives an increasing direction (because
$\beta_{1}>0$ and $\beta_{3}>0$). To find such a point, we use the
Lagrange multipliers method, as follows:

\begin{itemize}
\item Definition of the lagrangian\begin{eqnarray*}
\Lambda & = & \tau\ln2+H\ln H-H_{1}\ln\left(\frac{H_{1}}{e}\right)-H_{3}\ln\left(\frac{H_{3}}{e}\right)-H\\
 &  & +\sum_{\left(p,q\right)\in\mathcal{L}}d_{p,q}\ln d_{p,q}+\sum_{\substack{\left(p,q\right)\in\mathcal{L}\\
\left(i,0,k,l,0\right)\in Q_{p,q}\\
v\in\left\{ 0,1\right\} }
}\pi_{i,0,k,l,0,v}\ln\left(\omega_{i,0,k,l,0,v}\frac{{i+k \choose i,k}}{e\pi_{i,0,k,l,0,v}}\right)+1\\
 &  & -2c\ln3+2c\beta_{1}\ln\left(2\frac{\beta_{1}}{e}\right)+2c\beta_{3}\ln\left(3\frac{\beta_{3}}{e}\right)+2c\\
 &  & +\left(2c\ln b\right)\left(\beta_{1}+\beta_{3}-1\right)\\
 &  & +\sum_{\left(p,q\right)\in\mathcal{L}}\left(\ln r_{p,q}\right)\left(\sum_{\substack{\left(i,0,k,l,0\right)\in Q_{p,q}}
}\pi_{i,j,k,l,m,1}+\sum_{\substack{\left(i,0,k,l,0\right)\in Q_{q,p}}
}\pi_{i,j,k,l,m,0}-d_{p,q}\right)\\
 &  & +\left(\ln h\right)\left(H_{1}+H_{3}-H\right)\\
 &  & +\left(2\ln x_{1}\right)\left(\sum_{\substack{\left(p,q\right)\in\mathcal{L}\\
\left(i,0,k,l,0\right)\in Q_{p,q}\\
v\in\left\{ 0,1\right\} }
}i\pi_{i,0,k,l,0,v}+H_{1}-\beta_{1}c\right)+\left(\ln y_{1}\right)\left(\sum_{\substack{\left(p,q\right)\in\mathcal{L}\\
\left(i,0,k,l,0\right)\in Q_{p,q}\\
v\in\left\{ 0,1\right\} }
}l\pi_{i,0,k,l,0,v}-2\beta_{1}c\right)\end{eqnarray*}

\item Derivatives with respect to $\pi_{i,0,k,l,0,v}$
\end{itemize}
\begin{eqnarray*}
\frac{\partial\Lambda}{\partial\pi_{i,0,k,l,0,1}} & = & \ln\omega_{i,0,k,l,0,1}+\ln{i+k \choose i,k}-\ln\pi_{i,0,k,l,0,v}+\ln r_{i+k,l}+2i\ln x_{1}+l\ln y_{1}\\
\frac{\partial\Lambda}{\partial\pi_{i,0,k,l,0,0}} & = & \ln\omega_{i,0,k,l,0,0}+\ln{i+k \choose i,k}-\ln\pi_{i,0,k,l,0,v}+\ln r_{l,i+k}+2i\ln x_{1}+l\ln y_{1}\end{eqnarray*}

\begin{eqnarray*}
\pi_{i,0,k,l,0,1} & = & \omega_{i,0,k,l,0,1}{i+k \choose i,k}r_{i+k,l}x_{1}^{2i}y_{1}^{l}\\
\pi_{i,0,k,l,0,0} & = & \omega_{i,0,k,l,0,0}{i+k \choose i,k}r_{l,i+k}x_{1}^{2i}y_{1}^{l}\end{eqnarray*}

The $r_{p,q}$ contraints become:\begin{eqnarray*}
\sum_{\substack{\left(i,0,k,q,0\right)\in Q_{p,q}}
}\omega_{i,0,k,q,0,1}{i+k \choose i,k}r_{i+k,q}x_{1}^{2i}y_{1}^{q}\\
+\sum_{\substack{\left(i,0,k,p,0\right)\in Q_{q,p}}
}\omega_{i,0,k,p,0,0}{i+k \choose i,k}r_{l,i+k}x_{1}^{2i}y_{1}^{p} & = & d_{p,q}\end{eqnarray*}

Let us introduce $A_{p,q}=\sum_{\substack{\left(i,0,k,q,0\right)\in Q_{p,q}}
}\omega_{i,0,k,q,0,1}{p \choose i,k}x_{1}^{2i}y_{1}^{q}+\sum_{\substack{\left(i,0,k,l,0\right)\in Q_{q,p}}
}\omega_{i,0,k,p,0,0}{q \choose i,k}x_{1}^{2i}y_{1}^{p}$:\begin{eqnarray*}
r_{p,q}A_{p,q} & = & d_{p,q}\end{eqnarray*}
\begin{eqnarray*}
r_{p,q} & = & \frac{d_{p,q}}{A_{p,q}}\end{eqnarray*}

Thus\begin{eqnarray*}
\pi_{i,0,k,l,0,1} & = & \omega_{i,0,k,l,0,1}{i+k \choose i,k}\frac{d_{i+k,l}}{A_{i+k,l}}x_{1}^{2i}y_{1}^{l}\\
\pi_{i,0,k,l,0,0} & = & \omega_{i,0,k,l,0,0}{i+k \choose i,k}\frac{d_{l,i+k}}{A_{l,i+k}}x_{1}^{2i}y_{1}^{l}\end{eqnarray*}

\begin{itemize}
\item Derivatives with respect to $\beta_{t}$
\end{itemize}
\begin{eqnarray*}
\frac{\partial\Lambda}{\partial\beta_{1}} & = & 2c\ln2+2c\ln\beta_{1}+2c\ln b-2c\ln x_{1}-2c\ln y_{1}\\
\frac{\partial\Lambda}{\partial\beta_{3}} & = & 2c\ln3+2c\ln\beta_{3}+2c\ln b\end{eqnarray*}

\begin{eqnarray*}
\beta_{1} & = & \frac{x_{1}y_{1}}{2b}\\
\beta_{3} & = & \frac{1}{3b}\end{eqnarray*}

The $b$ constraint becomes:

\begin{eqnarray*}
\frac{x_{1}y_{1}}{2b}+\frac{1}{3b} & = & 1\end{eqnarray*}
\begin{eqnarray*}
b & = & \frac{x_{1}y_{1}}{2}+\frac{1}{3}\end{eqnarray*}

\begin{itemize}
\item Derivatives with respect to $H_{t}$
\end{itemize}
\begin{eqnarray*}
\frac{\partial\Lambda}{\partial H_{1}} & = & -\ln H_{1}+\ln h+2\ln x_{1}\\
\frac{\partial\Lambda}{\partial H_{3}} & = & -\ln H_{3}+\ln h\end{eqnarray*}

\begin{eqnarray*}
H_{1} & = & hx_{1}^{2}\\
H_{3} & = & h\end{eqnarray*}

The $h$ constraint becomes then:

\begin{eqnarray*}
h\left(x_{1}^{2}+1\right) & = & H\end{eqnarray*}

\begin{eqnarray*}
h & = & \frac{H}{x_{1}^{2}+1}\end{eqnarray*}

\begin{itemize}
\item Remaining constraints:
\end{itemize}
\begin{eqnarray*}
\sum_{\left(p,q\right)\in\mathcal{L}}\frac{d_{p,q}}{A_{p,q}}\sum_{\substack{\left(i,0,k,q,0\right)\in Q_{p,q}}
}i{p \choose i,k}x_{1}^{2i}y_{1}^{q}\omega_{i,0,k,q,0,1}\\
+\sum_{\left(p,q\right)\in\mathcal{L}}\frac{d_{p,q}}{A_{p,q}}\sum_{\substack{\left(i,0,k,p,0\right)\in Q_{q,p}}
}i{q \choose i,k}x_{1}^{2i}y_{1}^{q}\omega_{i,0,k,p,0,0}+\frac{Hx_{1}^{2}}{x_{1}^{2}+1} & = & \frac{x_{1}y_{1}c}{2\left(\frac{x_{1}y_{1}}{2}+\frac{1}{3}\right)}\\
\sum_{\left(p,q\right)\in\mathcal{L}}\frac{d_{p,q}}{A_{p,q}}\sum_{\substack{\left(i,0,k,q,0\right)\in Q_{p,q}}
}q{p \choose i,k}x_{1}^{2i}y_{1}^{q}\omega_{i,0,k,q,0,1}\\
+\sum_{\left(p,q\right)\in\mathcal{L}}\frac{d_{p,q}}{A_{p,q}}\sum_{\substack{\left(i,0,k,p,0\right)\in Q_{q,p}}
}p{q \choose i,k}x_{1}^{2i}y_{1}^{p}\omega_{i,0,k,p,0,0} & = & \frac{x_{1}y_{1}c}{\left(\frac{x_{1}y_{1}}{2}+\frac{1}{3}\right)}\end{eqnarray*}

Then we introduce $Z=\prod_{\left(p,q\right)\in\mathcal{L}}A_{p,q}^{d_{p,q}}$
and $Y=\ln Z$:\begin{eqnarray*}
\frac{\partial Y}{\partial x_{1}}+\frac{2Hx_{1}}{x_{1}^{2}+1} & = & \frac{y_{1}c}{\left(\frac{x_{1}y_{1}}{2}+\frac{1}{3}\right)}\\
\frac{\partial Y}{\partial y_{1}} & = & \frac{x_{1}c}{\left(\frac{x_{1}y_{1}}{2}+\frac{1}{3}\right)}\end{eqnarray*}

\end{itemize}
With \mathematica{} we found the following solutions (and the corresponding
values of $\ln F$):

\begin{table}
\caption{Interior point when $\beta_{2}=0$. }

\centering{}\begin{tabular}{|c|c|c|c|c|}
\hline 
model & standard & balanced signs & balanced occurrences & balanced signs and occurrences\tabularnewline
\hline
\hline 
$c$ & $4.500$ & $3.509$ & $4.623$ & $3.546$\tabularnewline
\hline 
$x_{1}$ & $0.997334$ & $0.96803$ & $0.999726$ & $0.970462$\tabularnewline
\hline 
$y_{1}$ & $2.07123$ & $2.06284$ & $2.05335$ & $2.06087$\tabularnewline
\hline 
$\ln F$ & $-2.463$ & $-1.78313$ & $-2.53084$ & $-1.79349$\tabularnewline
\hline
\end{tabular}
\end{table}

So all unforced variables are non zero.
\begin{itemize}
\item subcase where all unforced variables are non zero: we define a function
$f\left(\xi\right)$ representing the variation of $\ln F$ under
a small positive variation $\xi$ in the following direction; remember
that $\omega_{i,j,k,l,m,v}=1$ as soon as $i\geq1$ thus the corresponding
variable $\pi_{i,j,k,l,m,v}$ exists. Let us take some $p\geq3$ and
$q\geq2$ such that $d_{p,q}>0$. We make the following move: \begin{eqnarray*}
\beta_{1} & \to & \beta_{1}-\frac{\xi}{c}\\
\beta_{2} & \to & \beta_{2}+\frac{2\xi}{c}\\
\beta_{3} & \to & \beta_{3}-\frac{\xi}{c}\\
\pi_{p-1,0,1,q,0,1} & \to & \pi_{p-1,0,1,q,0,1}-5\xi\\
\pi_{p-2,1,1,q,0,1} & \to & \pi_{p-2,1,1,q,0,1}+\xi\\
\pi_{p-1,0,1,q-2,2,1} & \to & \pi_{p-1,0,1,q-2,2,1}+\xi\\
\pi_{p-1,1,0,q,0,1} & \to & \pi_{p-1,1,0,q,0,1}+3\xi\end{eqnarray*}
 so that all constraints remain satisfied; in fact we are performing
a small move inside the polytope $\mathcal{P}$ and we would like
to show that along this direction $\ln F$ is increasing. Note that
$\beta_{2}=\pi_{p-2,1,1,q,0,1}=\pi_{p-1,0,1,q-2,2,1}=\pi_{p-1,1,0,q,0,1}=0$
and all other variables here are non zero, so we have: \begin{eqnarray*}
f\left(\xi\right) & = & 2c\left(\beta_{1}-\frac{\xi}{c}\right)\ln\left(\frac{2}{e}\left(\beta_{1}-\frac{\xi}{c}\right)\right)-2c\beta_{1}\ln\left(\frac{2}{e}\beta_{1}\right)\\
 &  & +2c\left(\beta_{2}+\frac{2\xi}{c}\right)\ln\left(\frac{2}{e}\left(\beta_{2}+\frac{2\xi}{c}\right)\right)\\
 &  & +2c\left(\beta_{3}-\frac{\xi}{c}\right)\ln\left(\frac{3}{e}\left(\beta_{3}-\frac{\xi}{c}\right)\right)-2c\beta_{3}\ln\left(\frac{3}{e}\beta_{3}\right)\\
 &  & -\left(\pi_{p-1,0,1,q,0,1}-5\xi\right)\ln\left(\pi_{p-1,0,1,q,0,1}-5\xi\right)+\pi_{p-1,0,1,q,0,1}\ln\pi_{p-1,0,1,q,0,1}\\
 &  & -\left(\pi_{p-2,1,1,q,0,1}+\xi\right)\ln\left(\pi_{p-2,1,1,q,0,1}+\xi\right)-\left(\pi_{p-1,0,1,q-2,2,1}+\xi\right)\ln\left(\pi_{p-1,0,1,q-2,2,1}+\xi\right)\\
 &  & -\left(\pi_{p-1,1,0,q,0,1}+3\xi\right)\ln\left(\pi_{p-1,1,0,q,0,1}+3\xi\right)+\Theta\left(\xi\right)\end{eqnarray*}
\begin{eqnarray*}
f\left(\xi\right) & = & +2c\left(\beta_{1}-\frac{\xi}{c}\right)\ln\left(\frac{2}{e}\left(\beta_{1}-\frac{\xi}{c}\right)\right)-2c\beta_{1}\ln\left(\frac{2}{e}\beta_{1}\right)\\
 &  & +4\xi\ln\left(\frac{4\xi}{ec}\right)\\
 &  & +2c\left(\beta_{3}-\frac{\xi}{c}\right)\ln\left(\frac{3}{e}\left(\beta_{3}-\frac{\xi}{c}\right)\right)-2c\beta_{3}\ln\left(\frac{3}{e}\beta_{3}\right)\\
 &  & -\left(\pi_{p-1,0,1,q,0,1}-5\xi\right)\ln\left(\pi_{p-1,0,1,q,0,1}-5\xi\right)+\pi_{p-1,0,1,q,0,1}\ln\pi_{p-1,0,1,q,0,1}\\
 &  & -2\xi\ln\xi-3\xi\ln\left(3\xi\right)+\Theta\left(\xi\right)\end{eqnarray*}
and thus:\begin{eqnarray*}
\lim_{\xi\to0}\frac{f\left(\xi\right)}{\xi} & = & \lim_{\xi\to0}\left(-\ln\xi\right)+\Theta\left(1\right)\end{eqnarray*}

Since $\lim_{\xi\to0}\frac{f\left(\xi\right)}{\xi}=+\infty$, we have
found an increasing direction.

\end{itemize}
\item case where $\beta_{3}=0$: then $H_{3}=0$ and $\pi_{i,j,k,l,m,v}=0$
unless $k=0$; again we call these variables \emph{forced}.

\begin{itemize}
\item subcase where there is an unforced variable at zero: we find a feasible
point where $\beta_{3}=0$ and all unforced variables are non zero.
Then a move towards this point gives an increasing direction (because
$\beta_{1}>0$ and $\beta_{2}>0$). To find such a point we use again
the Lagrange multipliers method, as follows:

\begin{itemize}
\item Definition of the lagrangian
\end{itemize}
\begin{eqnarray*}
\Lambda & = & \tau\ln2+H\ln H-H_{1}\ln\left(\frac{H_{1}}{e}\right)-H_{2}\ln\left(\frac{H_{2}}{e}\right)-H\\
 &  & +\sum_{\left(p,q\right)\in\mathcal{L}}d_{p,q}\ln d_{p,q}+\sum_{\substack{\left(p,q\right)\in\mathcal{L}\\
\left(i,j,0,l,m\right)\in Q_{p,q}\\
v\in\left\{ 0,1\right\} }
}\pi_{i,j,0,l,m,v}\ln\left(\omega_{i,j,0,l,m,v}\frac{{i+j \choose i,j}{l+m \choose l,m}}{e\pi_{i,j,0,l,m,v}}\right)+1\\
 &  & -2c\ln3+2c\beta_{1}\ln\left(2\frac{\beta_{1}}{e}\right)+2c\beta_{2}\ln\left(2\frac{\beta_{2}}{e}\right)+2c\\
 &  & +\left(2c\ln b\right)\left(\beta_{1}+\beta_{2}-1\right)\\
 &  & +\sum_{\left(p,q\right)\in\mathcal{L}}\left(\ln r_{p,q}\right)\left(\sum_{\substack{\left(i,j,0,l,m\right)\in Q_{p,q}}
}\pi_{i,j,0,l,m,1}+\sum_{\substack{\left(i,j,0,l,m\right)\in Q_{q,p}}
}\pi_{i,j,0,l,m,0}-d_{p,q}\right)\\
 &  & +\left(\ln h\right)\left(H_{1}+H_{2}-H\right)\\
 &  & +\left(2\ln x_{1}\right)\left(\sum_{\substack{\left(p,q\right)\in\mathcal{L}\\
\left(i,j,0,l,m\right)\in Q_{p,q}\\
v\in\left\{ 0,1\right\} }
}i\pi_{i,j,0,l,m,v}+H_{1}-\beta_{1}c\right)+\left(\ln x_{2}\right)\left(\sum_{\substack{\left(p,q\right)\in\mathcal{L}\\
\left(i,j,0,l,m\right)\in Q_{p,q}\\
v\in\left\{ 0,1\right\} }
}j\pi_{i,j,0,l,m,v}+H_{2}-2\beta_{2}c\right)\\
 &  & +\left(\ln y_{1}\right)\left(\sum_{\substack{\left(p,q\right)\in\mathcal{L}\\
\left(i,j,0,l,m\right)\in Q_{p,q}\\
v\in\left\{ 0,1\right\} }
}l\pi_{i,j,0,l,m,v}-2\beta_{1}c\right)+\left(2\ln y_{2}\right)\left(\sum_{\substack{\left(p,q\right)\in\mathcal{L}\\
\left(i,j,0,l,m\right)\in Q_{p,q}\\
v\in\left\{ 0,1\right\} }
}m\pi_{i,j,0,l,m,v}-\beta_{2}c\right)\end{eqnarray*}

\begin{itemize}
\item Derivatives with respect to $\pi_{i,j,0,l,m,v}$
\end{itemize}
\begin{eqnarray*}
\frac{\partial\Lambda}{\partial\pi_{i,j,0,l,m,1}} & = & \ln\omega_{i,j,0,l,m,1}+\ln\left({i+j \choose i,j}{l+m \choose l,m}\right)-\ln\pi_{i,j,0,l,m,v}+\ln r_{i+j,l+m}\\
 &  & +2i\ln x_{1}+j\ln x_{2}+l\ln y_{1}+2m\ln y_{2}\\
\frac{\partial\Lambda}{\partial\pi_{i,j,0,l,m,0}} & = & \ln\omega_{i,j,0,l,m,0}+\ln\left({i+j \choose i,j}{l+m \choose l,m}\right)-\ln\pi_{i,j,0,l,m,v}+\ln r_{l+m,i+j}\\
 &  & +2i\ln x_{1}+j\ln x_{2}+l\ln y_{1}+2m\ln y_{2}\end{eqnarray*}

\begin{eqnarray*}
\pi_{i,j,0,l,m,1} & = & \omega_{i,j,0,l,m,1}{i+j \choose i,j}{l+m \choose l,m}r_{i+j,l+m}x_{1}^{2i}x_{2}^{j}y_{1}^{l}y_{2}^{2m}\\
\pi_{i,j,0,l,m,0} & = & \omega_{i,j,0,l,m,0}{i+j \choose i,j}{l+m \choose l,m}r_{l+m,i+j}x_{1}^{2i}x_{2}^{j}y_{1}^{l}y_{2}^{2m}\end{eqnarray*}

The $r_{p,q}$ contraints become:\begin{eqnarray*}
\sum_{\substack{\left(i,j,0,l,m\right)\in Q_{p,q}}
}\omega_{i,j,0,l,m,1}{i+j \choose i,j}{l+m \choose l,m}r_{i+j,l+m}x_{1}^{2i}x_{2}^{j}y_{1}^{l}y_{2}^{2m}\\
+\sum_{\substack{\left(i,j,0,l,m\right)\in Q_{q,p}}
}\omega_{i,j,0,l,m,0}{i+j \choose i,j}{l+m \choose l,m}r_{l+m,i+j}x_{1}^{2i}x_{2}^{j}y_{1}^{l}y_{2}^{2m} & = & d_{p,q}\end{eqnarray*}

Let us introduce \[
A_{p,q}=\sum_{\substack{\left(i,j,0,l,m\right)\in Q_{p,q}}
}\omega_{i,j,0,l,m,1}{p \choose i,j}{q \choose l,m}x_{1}^{2i}x_{2}^{j}y_{1}^{l}y_{2}^{2m}+\sum_{\substack{\left(i,j,0,l,m\right)\in Q_{q,p}}
}\omega_{i,j,0,l,m,0}{q \choose i,j}{p \choose l,m}x_{1}^{2i}x_{2}^{j}y_{1}^{l}y_{2}^{2m}\]
We have:\begin{eqnarray*}
r_{p,q}A_{p,q} & = & d_{p,q}\end{eqnarray*}
\begin{eqnarray*}
r_{p,q} & = & \frac{d_{p,q}}{A_{p,q}}\end{eqnarray*}

Thus\begin{eqnarray*}
\pi_{i,j,0,l,m,1} & = & \omega_{i,j,0,l,m,1}{i+j \choose i,j}{l+m \choose l,m}\frac{d_{i+j,l+m}}{A_{i+j,l+m}}x_{1}^{2i}x_{2}^{j}y_{1}^{l}y_{2}^{2m}\\
\pi_{i,j,0,l,m,0} & = & \omega_{i,j,0,l,m,0}{i+j \choose i,j}{l+m \choose l,m}\frac{d_{l+m,i+j}}{A_{l+m,i+j}}x_{1}^{2i}x_{2}^{j}y_{1}^{l}y_{2}^{2m}\end{eqnarray*}

\begin{itemize}
\item Derivatives with respect to $\beta_{t}$
\end{itemize}
\begin{eqnarray*}
\frac{\partial\Lambda}{\partial\beta_{1}} & = & 2c\ln2+2c\ln\beta_{1}+2c\ln b-2c\ln x_{1}-2c\ln y_{1}\\
\frac{\partial\Lambda}{\partial\beta_{2}} & = & 2c\ln2+2c\ln\beta_{2}+2c\ln b-2c\ln x_{2}-2c\ln y_{2}\end{eqnarray*}

\begin{eqnarray*}
\beta_{1} & = & \frac{x_{1}y_{1}}{2b}\\
\beta_{2} & = & \frac{x_{2}y_{2}}{2b}\end{eqnarray*}

The $b$ constraint becomes:

\begin{eqnarray*}
\frac{x_{1}y_{1}}{2b}+\frac{x_{2}y_{2}}{2b} & = & 1\end{eqnarray*}
\begin{eqnarray*}
b & = & \frac{x_{1}y_{1}}{2}+\frac{x_{2}y_{2}}{2}\end{eqnarray*}

\begin{itemize}
\item Derivatives with respect to $H_{t}$
\end{itemize}
\begin{eqnarray*}
\frac{\partial\Lambda}{\partial H_{1}} & = & -\ln H_{1}+\ln h+2\ln x_{1}\\
\frac{\partial\Lambda}{\partial H_{2}} & = & -\ln H_{2}+\ln h+\ln x_{2}\end{eqnarray*}

\begin{eqnarray*}
H_{1} & = & hx_{1}^{2}\\
H_{2} & = & hx_{2}\end{eqnarray*}

The $h$ constraint becomes then:

\begin{eqnarray*}
h\left(x_{1}^{2}+x_{2}\right) & = & H\end{eqnarray*}

\begin{eqnarray*}
h & = & \frac{H}{x_{1}^{2}+x_{2}}\end{eqnarray*}

\begin{itemize}
\item Remaining constraints:
\end{itemize}
\begin{eqnarray*}
\sum_{\left(p,q\right)\in\mathcal{L}}\frac{d_{p,q}}{A_{p,q}}\sum_{\substack{\left(i,j,0,l,m\right)\in Q_{p,q}}
}i{p \choose i,j}{q \choose l,m}x_{1}^{2i}x_{2}^{j}y_{1}^{l}y_{2}^{2m}\omega_{i,j,0,l,m,1}\\
+\sum_{\left(p,q\right)\in\mathcal{L}}\frac{d_{p,q}}{A_{p,q}}\sum_{\substack{\left(i,j,0,l,m\right)\in Q_{q,p}}
}i{q \choose i,j}{p \choose l,m}x_{1}^{2i}x_{2}^{j}y_{1}^{l}y_{2}^{2m}\omega_{i,j,0,l,m,0}+\frac{Hx_{1}^{2}}{x_{1}^{2}+x_{2}} & = & \frac{x_{1}y_{1}c}{\left(x_{1}y_{1}+x_{2}y_{2}\right)}\\
\sum_{\left(p,q\right)\in\mathcal{L}}\frac{d_{p,q}}{A_{p,q}}\sum_{\substack{\left(i,j,0,l,m\right)\in Q_{p,q}}
}j{p \choose i,j}{q \choose l,m}x_{1}^{2i}x_{2}^{j}y_{1}^{l}y_{2}^{2m}\omega_{i,j,0,l,m,1}\\
+\sum_{\left(p,q\right)\in\mathcal{L}}\frac{d_{p,q}}{A_{p,q}}\sum_{\substack{\left(i,j,0,l,m\right)\in Q_{q,p}}
}j{q \choose i,j}{p \choose l,m}x_{1}^{2i}x_{2}^{j}y_{1}^{l}y_{2}^{2m}\omega_{i,j,0,l,m,0}+\frac{Hx_{2}}{x_{1}^{2}+x_{2}} & = & \frac{2x_{2}y_{2}c}{\left(x_{1}y_{1}+x_{2}y_{2}\right)}\\
\sum_{\left(p,q\right)\in\mathcal{L}}\frac{d_{p,q}}{A_{p,q}}\sum_{\substack{\left(i,j,0,l,m\right)\in Q_{p,q}}
}l{p \choose i,j}{q \choose l,m}x_{1}^{2i}x_{2}^{j}y_{1}^{l}y_{2}^{2m}\omega_{i,j,0,l,m,1}\\
+\sum_{\left(p,q\right)\in\mathcal{L}}\frac{d_{p,q}}{A_{p,q}}\sum_{\substack{\left(i,j,0,l,m\right)\in Q_{q,p}}
}l{q \choose i,j}{p \choose l,m}x_{1}^{2i}x_{2}^{j}y_{1}^{l}y_{2}^{2m}\omega_{i,j,0,l,m,0} & = & \frac{2x_{1}y_{1}c}{\left(x_{1}y_{1}+x_{2}y_{2}\right)}\\
\sum_{\left(p,q\right)\in\mathcal{L}}\frac{d_{p,q}}{A_{p,q}}\sum_{\substack{\left(i,j,0,l,m\right)\in Q_{p,q}}
}m{p \choose i,j}{q \choose l,m}x_{1}^{2i}x_{2}^{j}y_{1}^{l}y_{2}^{2m}\omega_{i,j,0,l,m,1}\\
+\sum_{\left(p,q\right)\in\mathcal{L}}\frac{d_{p,q}}{A_{p,q}}\sum_{\substack{\left(i,j,0,l,m\right)\in Q_{q,p}}
}m{q \choose i,j}{p \choose l,m}x_{1}^{2i}x_{2}^{j}y_{1}^{l}y_{2}^{2m}\omega_{i,j,0,l,m,0} & = & \frac{x_{2}y_{2}c}{\left(x_{1}y_{1}+x_{2}y_{2}\right)}\end{eqnarray*}

Then we introduce $Z=\prod_{\left(p,q\right)\in\mathcal{L}}A_{p,q}^{d_{p,q}}$
and $Y=\ln Z$:\begin{eqnarray*}
\frac{\partial Y}{\partial x_{1}}+\frac{2Hx_{1}}{x_{1}^{2}+x_{2}} & = & \frac{y_{1}c}{\left(\frac{x_{1}y_{1}}{2}+\frac{x_{2}y_{2}}{2}\right)}\\
\frac{\partial Y}{\partial x_{2}}+\frac{H}{x_{1}^{2}+x_{2}} & = & \frac{y_{2}c}{\left(\frac{x_{1}y_{1}}{2}+\frac{x_{2}y_{2}}{2}\right)}\\
\frac{\partial Y}{\partial y_{1}} & = & \frac{x_{1}c}{\left(\frac{x_{1}y_{1}}{2}+\frac{x_{2}y_{2}}{2}\right)}\\
\frac{\partial Y}{\partial y_{2}} & = & \frac{x_{2}c}{\left(\frac{x_{1}y_{1}}{2}+\frac{x_{2}y_{2}}{2}\right)}\end{eqnarray*}

\end{itemize}
With \mathematica{} we found the following solutions (and the corresponding
values of $\ln F$):

\begin{table}
\caption{Interior point when $\beta_{3}=0$.}

\centering{}\begin{tabular}{|c|c|c|c|c|}
\hline 
model & standard & balanced signs & balanced occurrences & balanced signs and occurrences\tabularnewline
\hline
\hline 
$c$ & $4.500$ & $3.509$ & $4.623$ & $3.546$\tabularnewline
\hline 
$x_{1}$ & $0.512382$ & $0.473066$ & $0.493115$ & $0.494614$\tabularnewline
\hline 
$x_{2}$ & $0.546014$ & $0.489687$ & $0.501307$ & $0.494614$\tabularnewline
\hline 
$y_{1}$ & $0.583465$ & $0.520769$ & $0.531045$ & $0.494614$\tabularnewline
\hline 
$y_{2}$ & $0.529328$ & $0.513117$ & $0.505216$ & $0.494614$\tabularnewline
\hline 
$\ln F$ & $-0.682149$ & $-0.375917$ & $-0.695819$ & $-0.33427$\tabularnewline
\hline
\end{tabular}
\end{table}

So all unforced variables are non zero.
\begin{itemize}
\item subcase where all unforced variables are non zero: we define a function
$f\left(\xi\right)$ representing the variation of $\ln F$ under
a small positive variation $\xi$ in the following direction; remember
that $\omega_{i,j,k,l,m,v}=1$ as soon as $i\geq1$ thus the corresponding
variable $\pi_{i,j,k,l,m,v}$ exists. Let us take some $p\geq3$ and
$q\geq2$ such that $d_{p,q}>0$. We make the following move:\begin{eqnarray*}
\beta_{1} & \to & \beta_{1}+\frac{\xi}{c}\\
\beta_{2} & \to & \beta_{2}-\frac{2\xi}{c}\\
\beta_{3} & \to & \beta_{3}+\frac{\xi}{c}\\
\pi_{p-2,2,0,q-1,1,1} & \to & \pi_{p-2,2,0,q-1,1,1}-3\xi\\
\pi_{p-1,0,1,q-1,1,1} & \to & \pi_{p-1,0,1,q-1,1,1}+\xi\\
\pi_{p-2,1,1,q,0,1} & \to & \pi_{p-2,1,1,q,0,1}+2\xi\end{eqnarray*}
 so that all constraints remain satisfied; in fact we are performing
a small move inside the polytope $\mathcal{P}$ and we would like
to show that along this direction $\ln F$ is increasing. Note that
$\beta_{3}=\pi_{p-1,0,1,q-1,1,1}=\pi_{p-2,1,1,q,0,1}=0$ and all other
variables here are non zero, so we have:\begin{eqnarray*}
f\left(\xi\right) & = & 2c\left(\beta_{1}+\frac{\xi}{c}\right)\ln\left(\frac{2}{e}\left(\beta_{1}+\frac{\xi}{c}\right)\right)-2c\beta_{1}\ln\left(\frac{2}{e}\beta_{1}\right)\\
 &  & +2c\left(\beta_{2}-\frac{2\xi}{c}\right)\ln\left(\frac{2}{e}\left(\beta_{2}-\frac{2\xi}{c}\right)\right)-2c\beta_{2}\ln\left(\frac{2}{e}\beta_{2}\right)\\
 &  & +2c\left(\beta_{3}+\frac{\xi}{c}\right)\ln\left(\frac{3}{e}\left(\beta_{3}+\frac{\xi}{c}\right)\right)\\
 &  & -\left(\pi_{p-2,2,0,q-1,1,1}-3\xi\right)\ln\left(\pi_{p-2,2,0,q-1,1,1}-3\xi\right)+\pi_{p-2,2,0,q-1,1,1}\ln\pi_{p-2,2,0,q-1,1,1}\\
 &  & -\left(\pi_{p-1,0,1,q-1,1,1}+\xi\right)\ln\left(\pi_{p-1,0,1,q-1,1,1}+\xi\right)-\left(\pi_{p-2,1,1,q,0,1}+2\xi\right)\ln\left(\pi_{p-2,1,1,q,0,1}+2\xi\right)+\Theta\left(\xi\right)\end{eqnarray*}
\begin{eqnarray*}
f\left(\xi\right) & = & +2c\left(\beta_{1}+\frac{\xi}{c}\right)\ln\left(\frac{2}{e}\left(\beta_{1}+\frac{\xi}{c}\right)\right)-2c\beta_{1}\ln\left(\frac{2}{e}\beta_{1}\right)\\
 &  & +2c\left(\beta_{2}-\frac{2\xi}{c}\right)\ln\left(\frac{2}{e}\left(\beta_{2}-\frac{2\xi}{c}\right)\right)-2c\beta_{2}\ln\left(\frac{2}{e}\beta_{2}\right)\\
 &  & +2\xi\ln\left(\frac{3\xi}{ec}\right)\\
 &  & -\left(\pi_{p-2,2,0,q-1,1,1}-3\xi\right)\ln\left(\pi_{p-2,2,0,q-1,1,1}-3\xi\right)+\pi_{p-2,2,0,q-1,1,1}\ln\pi_{p-2,2,0,q-1,1,1}\\
 &  & -\xi\ln\xi-2\xi\ln\left(2\xi\right)+\Theta\left(\xi\right)\end{eqnarray*}
and thus:\begin{eqnarray*}
\lim_{\xi\to0}\frac{f\left(\xi\right)}{\xi} & = & \lim_{\xi\to0}\left(-\ln\xi\right)+\Theta\left(1\right)\end{eqnarray*}

Since $\lim_{\xi\to0}\frac{f\left(\xi\right)}{\xi}=+\infty$, we have
found an increasing direction.

\end{itemize}
\item case where all $\beta_{t}>0$; suppose there is another variable at
zero; we move towards the general solution we found in appendix \ref{sec:General-Lagrange},
where all variables are non zero. Then again $\lim_{\xi\to0}\frac{f\left(\xi\right)}{\xi}=+\infty$;
so this is an increasing direction.
\end{enumerate}

\section{Inspection of the Interior of $\mathcal{P}$\label{sec:Inspection-interior}}

As \cite{Diaz2009} noticed in their calculation, we can perform a
sweep over some coordinates in order to check that the solution of
the Lagrange multipliers problem is indeed a global maximum. Namely
when we fix all $\beta_{1}$, $\beta_{2}$ (and $\beta_{3}=1-\beta_{1}-\beta_{2}$),
the function $\ln F$ is strictly concave in the other variables.
Let $\mathcal{P}_{\beta_{1},\beta_{2}}$ the polytope where the remaining
variables are allowed to move ; remember that the function to maximize
is:

\begin{eqnarray*}
\ln F & = & \tau\ln2+H\ln H-H_{1}\ln\left(\frac{H_{1}}{e}\right)-H_{2}\ln\left(\frac{H_{2}}{e}\right)-H_{3}\ln\left(\frac{H_{3}}{e}\right)-H\\
 &  & +\sum_{\left(p,q\right)\in\mathcal{L}}d_{p,q}\ln d_{p,q}+\sum_{\substack{\left(p,q\right)\in\mathcal{L}\\
\left(i,j,k,l,m\right)\in Q_{p,q}\\
v\in\left\{ 0,1\right\} }
}\pi_{i,j,k,l,m,v}\ln\left(\omega_{i,j,k,l,m,v}\frac{{i+j+k \choose i,j,k}{l+m \choose l,m}}{e\pi_{i,j,k,l,m,v}}\right)+1\\
 &  & -2c\ln3+2c\beta_{1}\ln\left(2\frac{\beta_{1}}{e}\right)+2c\beta_{2}\ln\left(2\frac{\beta_{2}}{e}\right)+2c\beta_{3}\ln\left(3\frac{\beta_{3}}{e}\right)+2c\end{eqnarray*}

If we increase an $H_{t}$ or a $\pi_{i,j,k,l,m,v}$ from $0$ to
a small $\xi>0$ and we change any other non zero variables, then
the variation $f$ of $\ln F$: $f=-\xi\ln\xi+\Theta\left(\xi\right)$
is such that $\frac{f}{\xi}=-\ln\xi+\Theta\left(1\right)\to+\infty$,
so $\ln F$ must increase; thus the function cannot maximize on the
boundary of $\mathcal{P}_{\beta_{1},\beta_{2}}$ and we can apply
the Lagrange multiplier technique again. But now by strict concavity
of the objective function, we know that the solution we find corresponds
to a global maximum.
\begin{itemize}
\item Definition of the lagrangian

\begin{eqnarray*}
\Lambda & = & \tau\ln2+H\ln H-H_{1}\ln\left(\frac{H_{1}}{e}\right)-H_{2}\ln\left(\frac{H_{2}}{e}\right)-H_{3}\ln\left(\frac{H_{3}}{e}\right)-H\\
 &  & +\sum_{\left(p,q\right)\in\mathcal{L}}d_{p,q}\ln d_{p,q}+\sum_{\substack{\left(p,q\right)\in\mathcal{L}\\
\left(i,j,k,l,m\right)\in Q_{p,q}\\
v\in\left\{ 0,1\right\} }
}\pi_{i,j,k,l,m,v}\ln\left(\omega_{i,j,k,l,m,v}\frac{{i+j+k \choose i,j,k}{l+m \choose l,m}}{e\pi_{i,j,k,l,m,v}}\right)+1\\
 &  & -2c\ln3+2c\beta_{1}\ln\left(2\frac{\beta_{1}}{e}\right)+2c\beta_{2}\ln\left(2\frac{\beta_{2}}{e}\right)+2c\beta_{3}\ln\left(3\frac{\beta_{3}}{e}\right)+2c\\
 &  & +\sum_{\left(p,q\right)\in\mathcal{L}}\left(\ln r_{p,q}\right)\left(\sum_{\substack{\left(i,j,k,l,m\right)\in Q_{p,q}}
}\pi_{i,j,k,l,m,1}+\sum_{\substack{\left(i,j,k,l,m\right)\in Q_{q,p}}
}\pi_{i,j,k,l,m,0}-d_{p,q}\right)\\
 &  & +\left(\ln h\right)\left(H_{1}+H_{2}+H_{3}-\sum_{\left(p,q\right)\in\mathcal{H}}\left(p+q\right)d_{p,q}\right)\\
 &  & +\left(2\ln x_{1}\right)\left(\sum_{\substack{\left(p,q\right)\in\mathcal{L}\\
\left(i,j,k,l,m\right)\in Q_{p,q}\\
v\in\left\{ 0,1\right\} }
}i\pi_{i,j,k,l,m,v}+H_{1}-\beta_{1}c\right)+\left(\ln x_{2}\right)\left(\sum_{\substack{\left(p,q\right)\in\mathcal{L}\\
\left(i,j,k,l,m\right)\in Q_{p,q}\\
v\in\left\{ 0,1\right\} }
}j\pi_{i,j,k,l,m,v}+H_{2}-2\beta_{2}c\right)\\
 &  & +\left(\ln y_{1}\right)\left(\sum_{\substack{\left(p,q\right)\in\mathcal{L}\\
\left(i,j,k,l,m\right)\in Q_{p,q}\\
v\in\left\{ 0,1\right\} }
}l\pi_{i,j,k,l,m,v}-2\beta_{1}c\right)+\left(2\ln y_{2}\right)\left(\sum_{\substack{\left(p,q\right)\in\mathcal{L}\\
\left(i,j,k,l,m\right)\in Q_{p,q}\\
v\in\left\{ 0,1\right\} }
}m\pi_{i,j,k,l,m,v}-\beta_{2}c\right)\end{eqnarray*}

\item Derivatives with respect to $\pi_{i,j,k,l,m,v}$

As in the general case we find that\begin{eqnarray*}
\pi_{i,j,k,l,m,1} & = & \omega_{i,j,k,l,m,1}{i+j+k \choose i,j,k}{l+m \choose l,m}\frac{d_{i+j+k,l+m}}{A_{i+j+k,l+m}}x_{1}^{2i}x_{2}^{j}y_{1}^{l}y_{2}^{2m}\\
\pi_{i,j,k,l,m,0} & = & \omega_{i,j,k,l,m,0}{i+j+k \choose i,j,k}{l+m \choose l,m}\frac{d_{l+m,i+j+k}}{A_{l+m,i+j+k}}x_{1}^{2i}x_{2}^{j}y_{1}^{l}y_{2}^{2m}\end{eqnarray*}

\item Derivatives with respect to $H_{t}$

As in the general case we find that

\begin{eqnarray*}
H_{1} & = & \frac{Hx_{1}^{2}}{x_{1}^{2}+x_{2}+1}\\
H_{2} & = & \frac{Hx_{2}}{x_{1}^{2}+x_{2}+1}\\
H_{3} & = & \frac{H}{x_{1}^{2}+x_{2}+1}\end{eqnarray*}

\item Remaining constraints:

\begin{eqnarray*}
\frac{\partial Y}{\partial x_{1}}+\frac{2Hx_{1}}{x_{1}^{2}+x_{2}+1} & = & \frac{2\beta_{1}c}{x_{1}}\\
\frac{\partial Y}{\partial x_{2}}+\frac{H}{x_{1}^{2}+x_{2}+1} & = & \frac{2\beta_{2}c}{x_{2}}\\
\frac{\partial Y}{\partial y_{1}} & = & \frac{2\beta_{1}c}{y_{1}}\\
\frac{\partial Y}{\partial y_{2}} & = & \frac{2\beta_{2}c}{y_{2}}\end{eqnarray*}

\item Objective function:

\begin{eqnarray*}
F & = & 2^{\tau}\frac{H^{H}}{H_{1}^{H_{1}}H_{2}^{H_{2}}H_{3}^{H_{3}}}\prod_{\left(p,q\right)\in\mathcal{L}}d_{p,q}^{d_{p,q}}\prod_{\substack{\left(p,q\right)\in\mathcal{L}\\
\left(i,j,k,l,m\right)\in Q_{p,q}\\
v\in\left\{ 0,1\right\} }
}\left(\omega_{i,j,k,l,m,v}\frac{{i+j+k \choose i,j,k}{l+m \choose l,m}}{\pi_{i,j,k,l,m,v}}\right)^{\pi_{i,j,k,l,m,v}}\left(\frac{1}{3}\left(2\beta_{1}\right)^{\beta_{1}}\left(2\beta_{2}\right)^{\beta_{2}}\left(3\beta_{3}\right)^{\beta_{3}}\right)^{2c}\end{eqnarray*}
\begin{eqnarray*}
F & = & 2^{\tau}\left(x_{1}^{2}+x_{2}+1\right)^{H}\prod_{\left(p,q\right)\in\mathcal{L}}\left(A_{p,q}^{d_{p,q}}\right)\left(\frac{1}{3}\left(\frac{2\beta_{1}}{x_{1}y_{1}}\right)^{\beta_{1}}\left(\frac{2\beta_{2}}{x_{2}y_{2}}\right)^{\beta_{2}}\left(3\beta_{3}\right)^{\beta_{3}}\right)^{2c}\end{eqnarray*}

\end{itemize}
So we made a sweep over $\beta_{1}$ and $\beta_{2}$ in the feasible
region and plotted the maximum point given as the solution of these
equations, which confirmed the fact that the solutions to the global
Lagrange system are indeed global maxima.

\begin{table}
\caption{Summary of results obtained by Lagrange's method.}

\centering{}\begin{tabular}{|>{\centering}p{0.12\paperwidth}|>{\centering}p{0.1\paperwidth}|>{\centering}p{0.1\paperwidth}|>{\centering}p{0.1\paperwidth}|>{\centering}p{0.13\paperwidth}|}
\hline 
model & standard & balanced signs & balanced occurrences & balanced signs and occurrences\tabularnewline
\hline 
our method

$\left(\alpha\rho_{j,l}+\rho_{k,m},v\right)>\left(0,0\right)$ & $4.500$ & $3.509$ & $4.623$ & $3.546$\tabularnewline
\hline 
$\beta_{1}$ & $0.44373$ & $0.557479$ & $0.445306$ & $0.568436$\tabularnewline
\hline 
$\beta_{2}$ & $0.421847$ & $0.365723$ & $0.421418$ & $0.363128$\tabularnewline
\hline 
$\beta_{3}$ & $0.134422$ & $0.0767974$ & $0.133276$ & $0.0684362$\tabularnewline
\hline
\end{tabular}
\end{table}

\begin{figure}
\begin{centering}
\includegraphics[width=0.45\columnwidth]{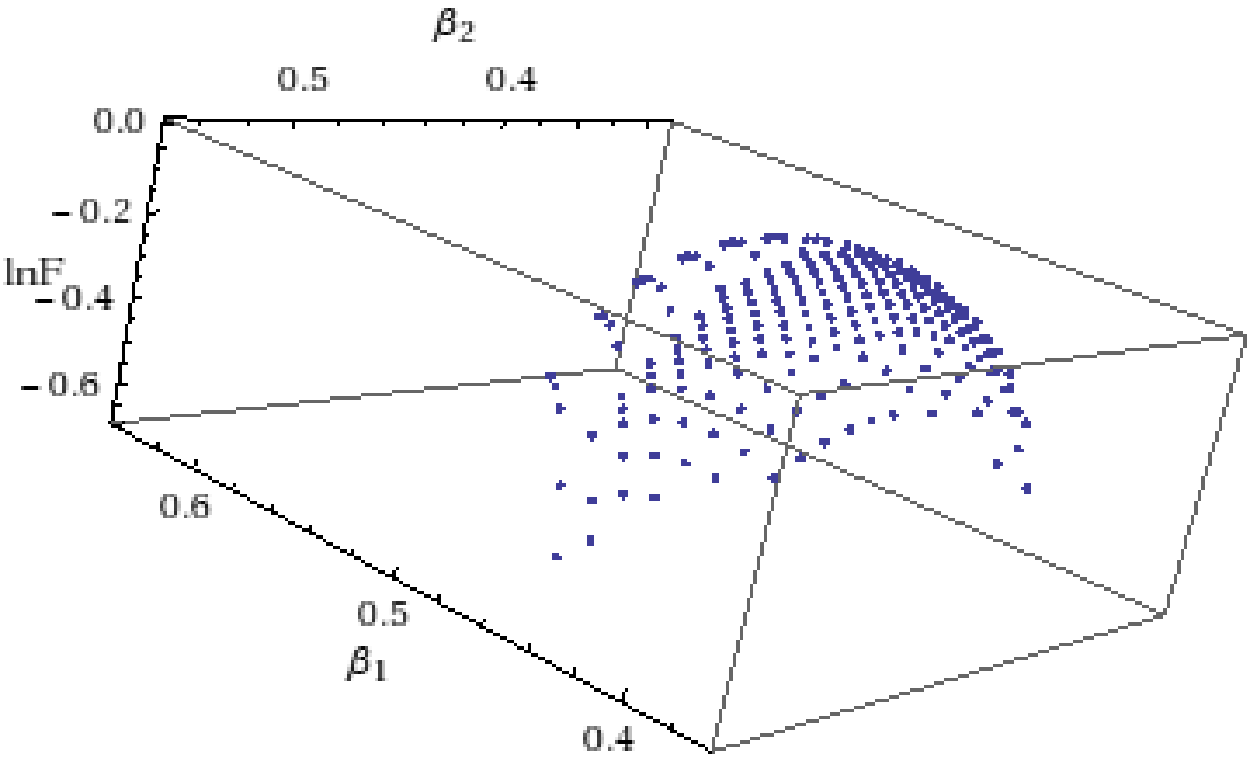}
\par\end{centering}

\includegraphics[width=0.45\columnwidth]{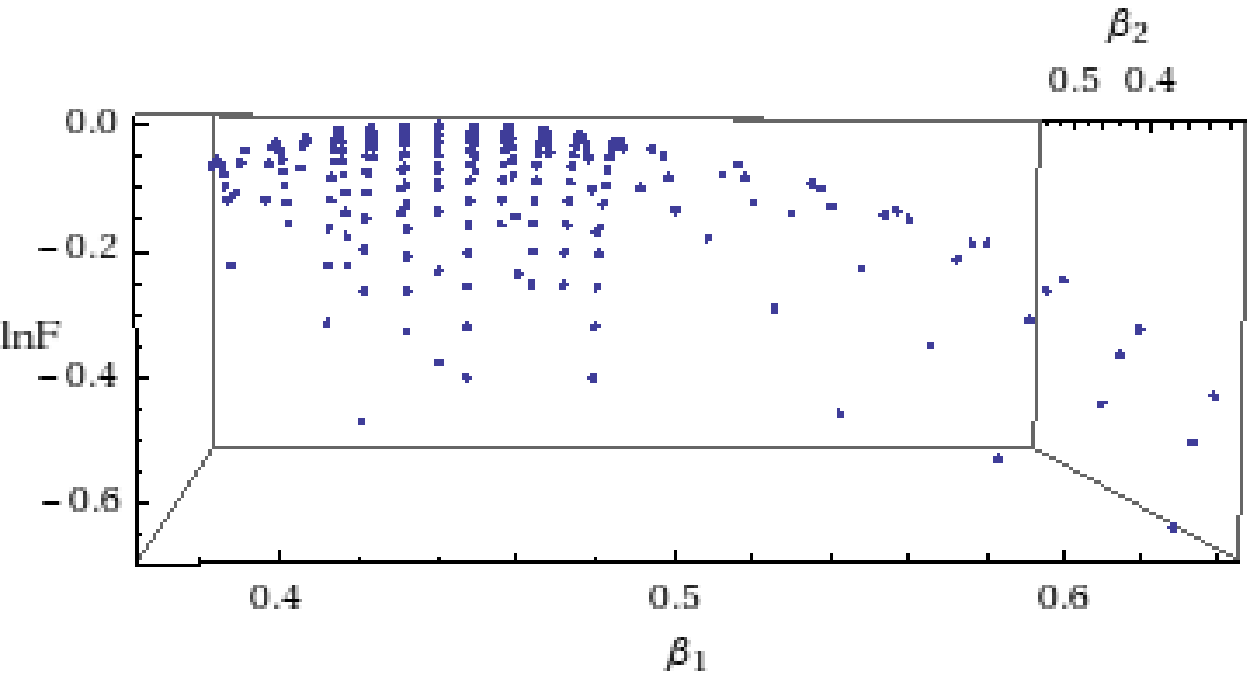}\hfill{}\includegraphics[width=0.45\columnwidth]{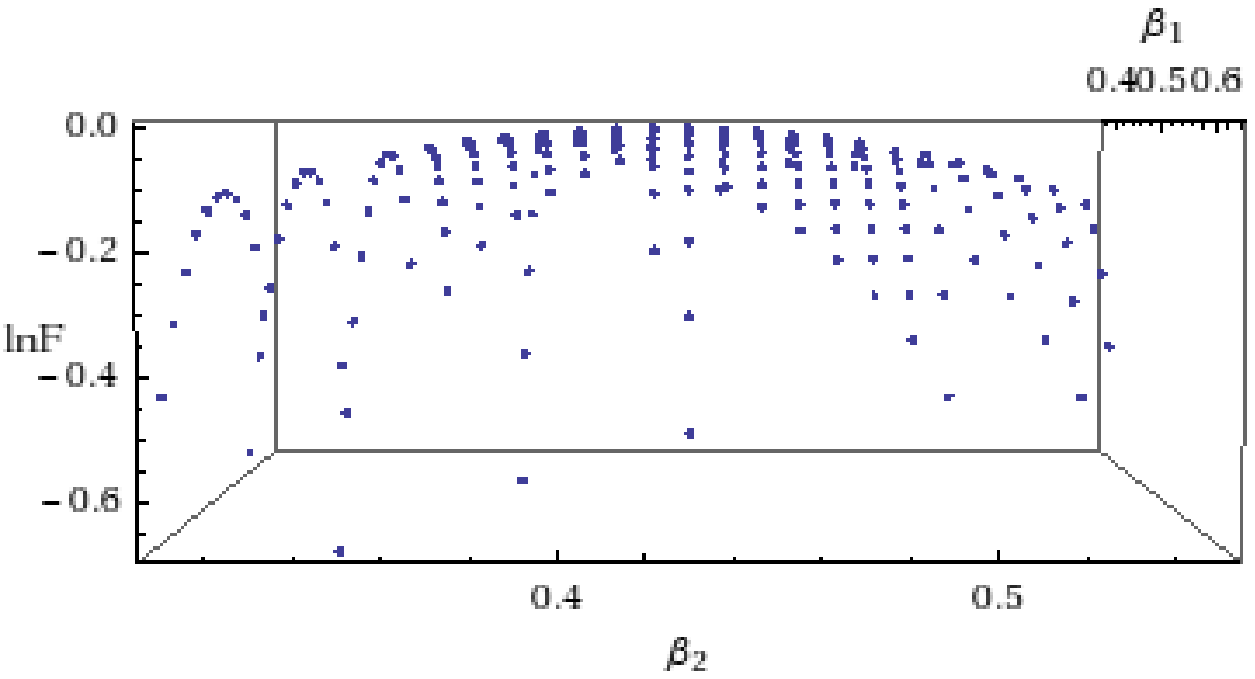}\caption{Maximum of $\ln F$ for different values of $\beta_{1}$ and $\beta_{2}$
in the standard model at $c=4.500$. Numerically we found that the
maximum is at $\beta_{1}\simeq0.44313$ and $\beta_{2}\simeq0.421847$.}

\end{figure}
\begin{figure}
\begin{centering}
\includegraphics[width=0.45\columnwidth]{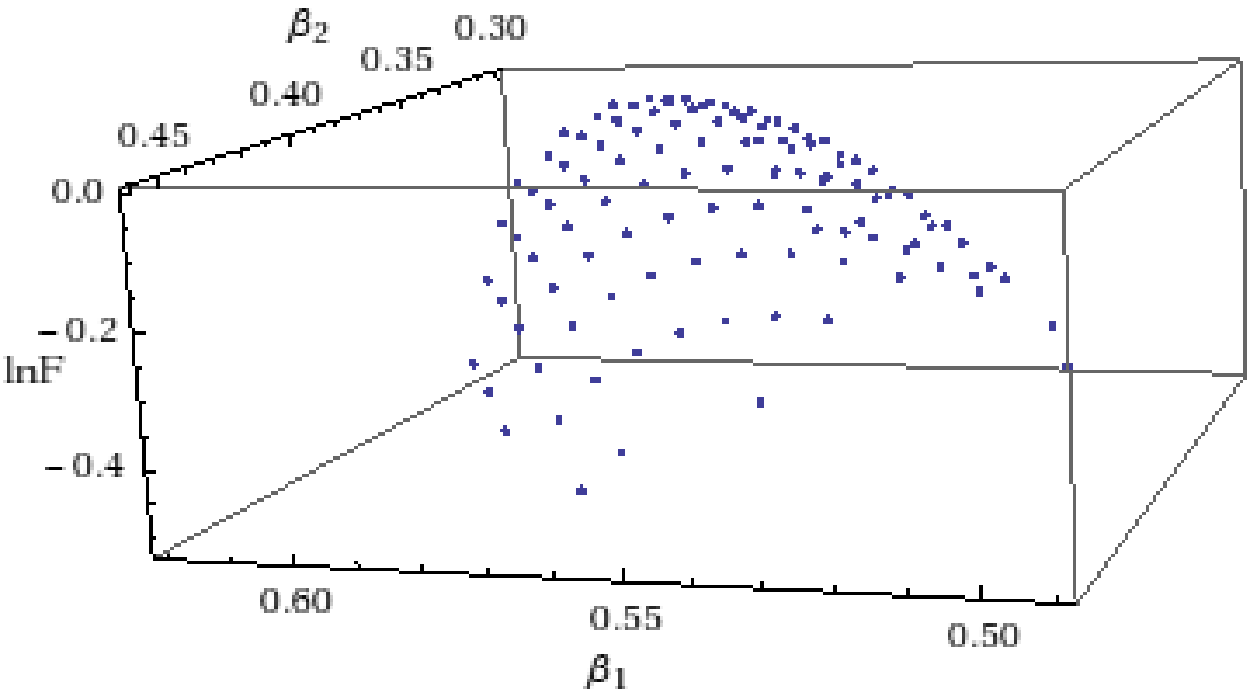}
\par\end{centering}

\includegraphics[width=0.45\columnwidth]{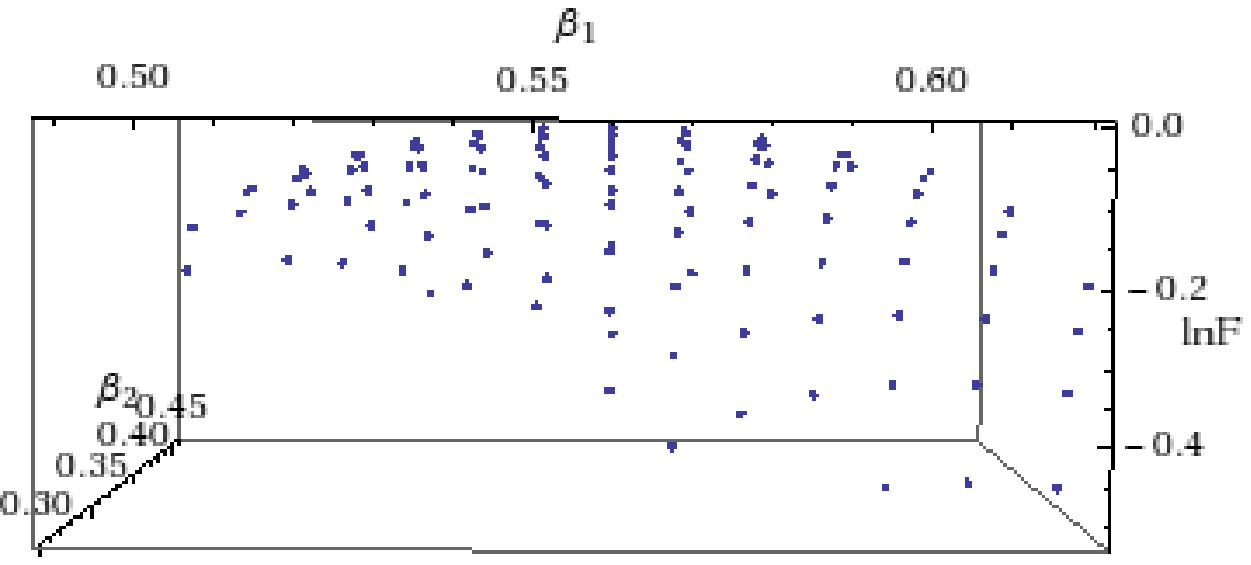}\hfill{}\includegraphics[width=0.45\columnwidth]{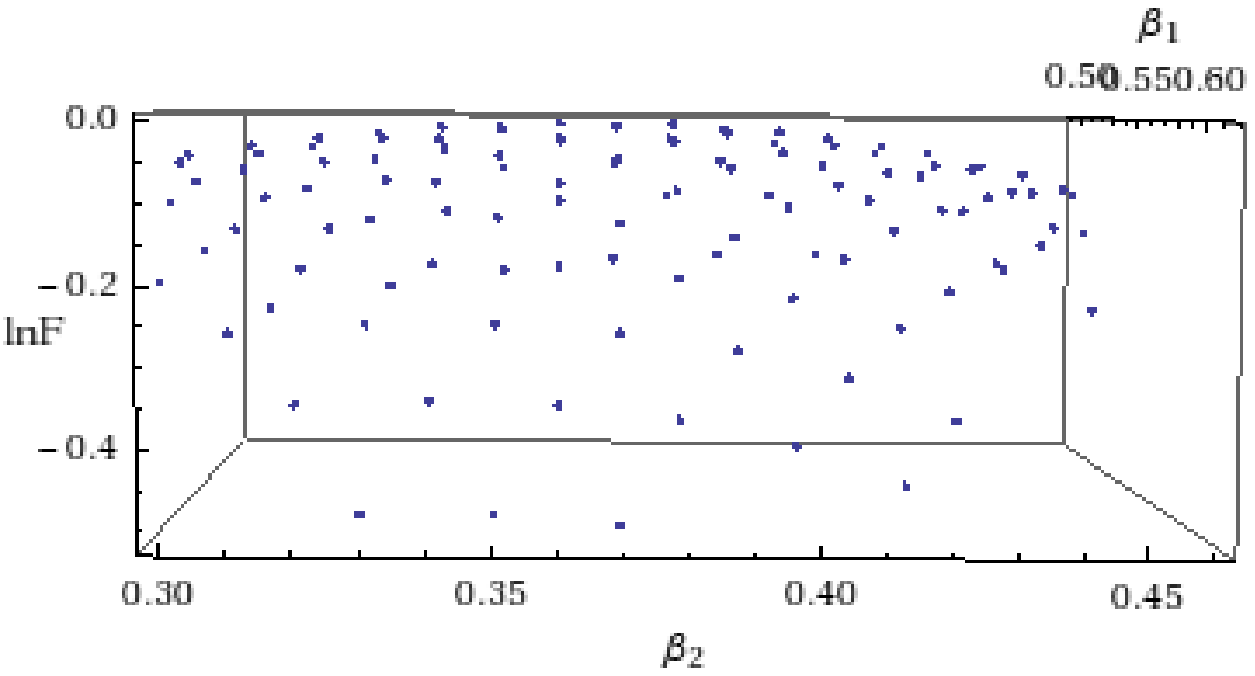}

\caption{Maximum of $\ln F$ for different values of $\beta_{1}$ and $\beta_{2}$
in the model with balanced signs at $c=3.509$. Numerically we found
that the maximum is at $\beta_{1}\simeq0.557479$ and $\beta_{2}\simeq0.365723$.}

\end{figure}
\begin{figure}
\begin{centering}
\includegraphics[width=0.45\columnwidth]{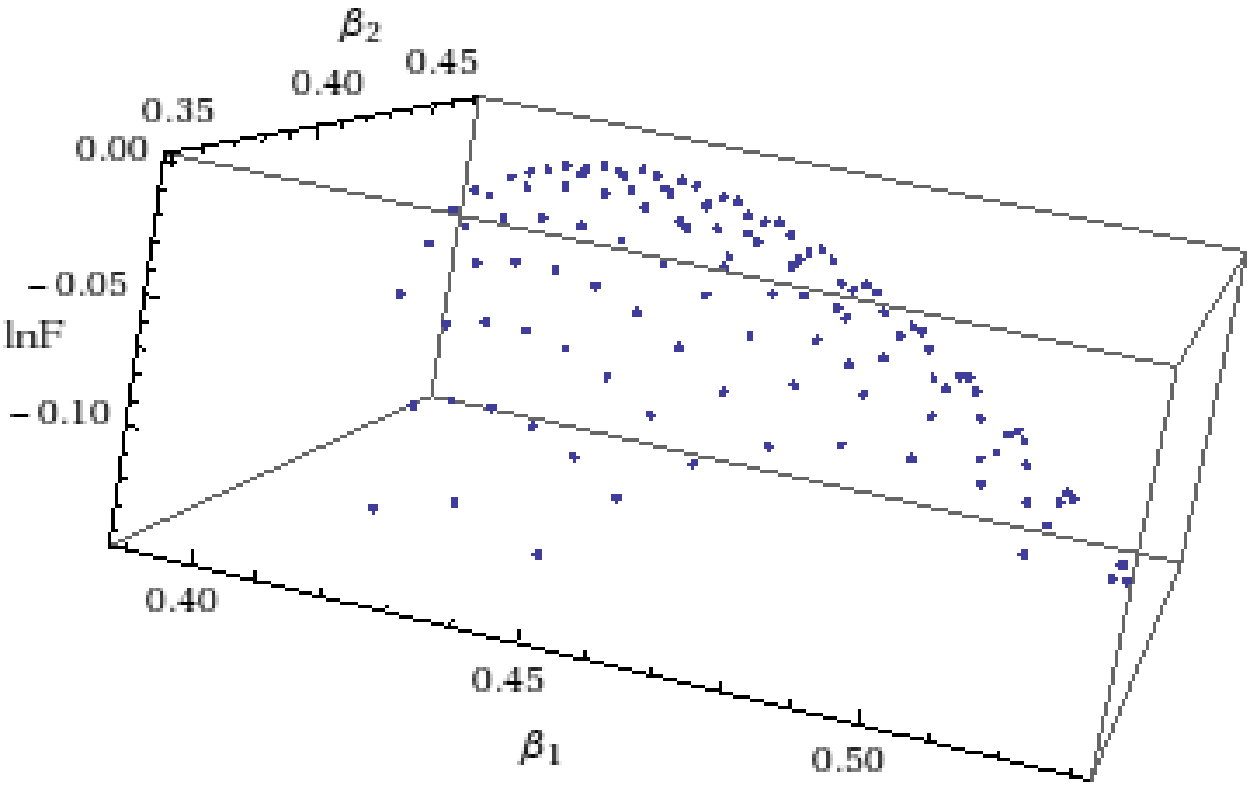}
\par\end{centering}

\includegraphics[width=0.45\columnwidth]{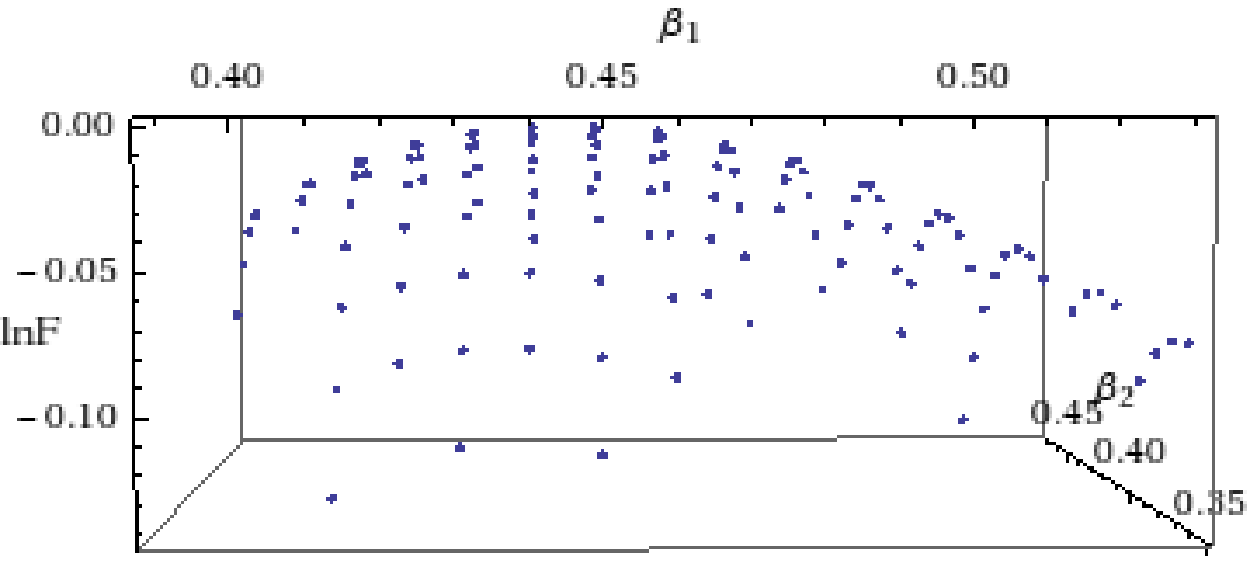}\hfill{}\includegraphics[width=0.45\columnwidth]{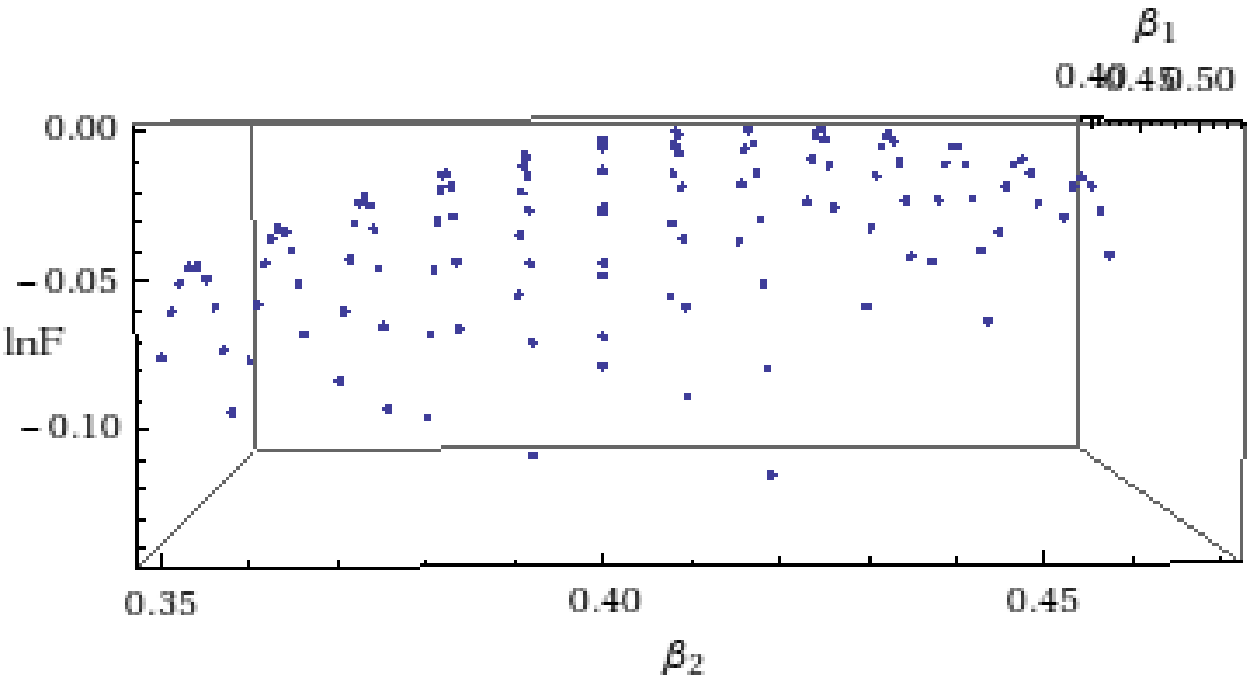}

\caption{Maximum of $\ln F$ for different values of $\beta_{1}$ and $\beta_{2}$
in the model with balanced occurrences at $c=4.623$. Numerically
we found that the maximum is at $\beta_{1}\simeq0.445306$ and $\beta_{2}\simeq0.421418$.}

\end{figure}
\begin{figure}
\begin{centering}
\includegraphics[width=0.5\paperwidth]{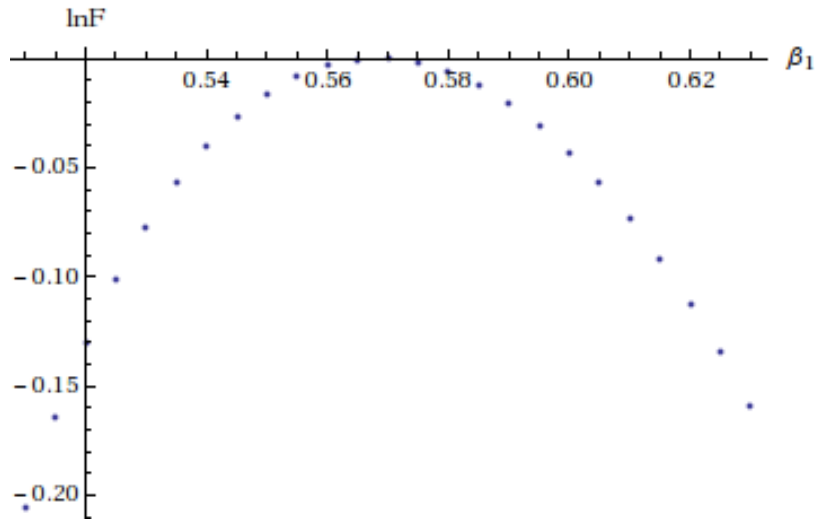}
\par\end{centering}

\caption{Maximum of $\ln F$ for different values of $\beta_{1}$ in the model
with balanced signs and occurrences at $c=3.546$. In this particular
model where each variable has as many positive occurrences as negative
ones, true and false surfaces are equal: $\beta_{1}+2\beta_{2}+3\beta_{3}=2\beta_{1}+\beta_{2}$,
thus $\beta_{2}=1.5-2\beta_{1}$. Numerically we found that the maximum
is at $\beta_{1}\simeq0.568436$ and $\beta_{2}\simeq0.363128$.}

\end{figure}

\end{document}